\documentstyle[onecolumn]{mn}
\input{psfig} 
\input{epsf.tex}
\newcommand{\mincir}{\raise -2.truept\hbox{\rlap
{\hbox{$\sim$}}\raise5.truept
\hbox{$<$}\ }}
\newcommand{\magcir}{\raise -2.truept\hbox{\rlap
{\hbox{$\sim$}}\raise5.truept
\hbox{$>$}\ }}
\newcommand{\minmag}{\raise-2.truept\hbox{\rlap
{\hbox{$<$}}\raise 6.truept\hbox
{$>$}\ }}
\newcommand{\be}{\begin{equation}}
\newcommand{\ee}{\end{equation}}
\newcommand{\ba}{\begin{eqnarray}}
\newcommand{\ea}{\end{eqnarray}}
\newcommand{\brr}{\begin{array}}
\newcommand{\err}{\end{array}}
\newcommand{\bc}{\begin{center}}
\newcommand{\ec}{\end{center}}

\newcommand{\bx}{\mbox{\bf x}}

\newcommand{\hm}{\,h^{-1}{\rm Mpc}}
\newcommand{\vel}{\,{\rm km\,s^{-1}}}
\newcommand{\om}{\omega}
\newcommand{\etal}{{et al.}~}

\newcommand{\p}{\partial}
\newcommand{\f}{\frac}

\newcommand{\w}{\omega}
\newcommand{\de}{\delta}
\newcommand{\ded}{\delta_{_D}}

\newcommand{\s}{\sigma}

\newcommand{\Lam}{\Lambda}
\newcommand{\fde}{\tilde{\delta}}

\newcommand{\fW}{\widetilde{W}}
\newcommand{\bfx}{{\bf x}}
\newcommand{\bfy}{{\bf y}}
\newcommand{\bfk}{{\bf k}}

\newcommand{\bfJ}{{\bf J}}

\newcommand{\bfr}{{\bf r}}

\newcommand{\calW}{{\cal W}}

\newcommand{\calP}{{\cal P}}

\newcommand{\calG}{{\cal G}}
\newcommand{\calF}{{\cal F}}
\newcommand{\lan}{\langle}
\newcommand{\ran}{\rangle}

\newcommand{\0}{\circ}

\title[Excursion set approach to halo clustering] 
{Excursion set approach to the clustering of dark matter haloes in 
Lagrangian space}
\author[C. Porciani, S. Matarrese, F. Lucchin \& P. Catelan]
{ { \Large Cristiano Porciani$^{1}$, Sabino Matarrese$^{2}$,
Francesco Lucchin$^{3}$ and Paolo Catelan$^{4,5}$} \\ 
$^1$ SISSA, Scuola Internazionale di Studi Superiori Avanzati,
via Beirut 2-4, I-34014 Trieste, Italy\\
$^2$ Dipartimento di Fisica {\em Galileo Galilei}, Universit\`{a} di
Padova, via Marzolo 8, I-35131 Padova, Italy\\
$^3$ Dipartimento di Astronomia, Universit\`{a} di Padova, vicolo
dell'Osservatorio 5, I-35122 Padova, Italy\\
$^4$ Theoretical Astrophysics Center, Juliane Maries Vej 30, DK-2100 
Copenhagen $\O$, Denmark\\
$^5$ Department of Physics, Astrophysics, Nuclear Physics Laboratory,
Keble Road OX1 3RH, Oxford, UK \\}

\begin{document}

\maketitle

\begin{abstract}
We present a stochastic approach to the spatial clustering of dark
matter haloes in Lagrangian space. Our formalism is based on a local
formulation of the `excursion set' approach by Bond et al., which
automatically accounts for the `cloud-in-cloud' problem in the
identification of bound systems. Our method allows to calculate
correlation functions of haloes in Lagrangian space using either a
multi-dimensional Fokker-Planck equation with suitable boundary
conditions or an array of Langevin equations with spatially correlated
random forces. We compare the results of our method with theoretical
predictions for the halo auto-correlation function considered in the
literature and find good agreement with the results recently obtained
within a treatment of halo clustering in terms of `counting fields' by
Catelan et al.. The possible effect of spatial correlations on
numerical simulations of halo merger trees is finally discussed.
\end{abstract}


\section{Introduction}

The celebrated Press \& Schechter (1974, hereafter PS) theory
represents a fundamental tool to determine the mean mass distribution
of bound condensations, nowadays identified as dark matter (DM) haloes,
in the Universe. Recently a number of authors (e.g. Efstathiou \etal
1988; Cole \& Kaiser 1989; Mo \& White 1996; Mo, Jing \& White 1996;
Catelan \etal 1997, hereafter CLMP) went beyond this application,
extending the PS method to study halo clustering both in Lagrangian and
Eulerian space.

The clustering problem can be dealt with in two steps: {\em i)}
identifying the preferential sites for halo formation in Lagrangian
space and {\em ii)} providing a theoretical framework to follow the
non-linear dynamics of the halo distribution. The former step can be
accurately modeled within linear theory by the PS scheme, which in fact
suggests a simple criterion to assign each Lagrangian point to a matter
clump of some mass $M$ which is going to collapse at some formation
redshift $z_f$. The latter step, instead, requires some knowledge of
the non-linear matter dynamics, which is ultimately responsible for the
actual Eulerian distribution of matter clumps as they form by the
hierarchical process of matter accretion and merging of subunits.  A
relevant progress in this direction has been made by Mo \& White
(1996), who showed that a linear local bias relation connects the
Eulerian halo auto-correlation function to that of the mass, thus
providing a good fit to the two-point function of haloes in N-body
simulations. A more refined, non-linear and non-local relation has been
recently obtained by CLMP.

At the Lagrangian level, however, one has to deal with the so-called
cloud-in-cloud problem, inherent in the PS scheme, namely the fact that
their approach selects bound systems of given mass that can have been
already included in larger mass condensations of the same catalogue. A
straightforward solution of this problem in connection with halo
correlations was sketched by CLMP, who adopted the `peak-background'
split (e.g. Bardeen \etal 1986; Cole \& Kaiser 1989) to solve the
problem up to the resolution scale of the `background' mass component,
which is also responsible for the motion of the haloes from their
original Lagrangian positions.

The transition from the Lagrangian to the Eulerian halo distribution
can be dealt with exactly thanks to the continuity equation, as shown
by CLMP, who obtained an analytic formula for the Eulerian `bias field'
connecting the halo number density fluctuations to the mass
perturbations. This relation possesses the remarkable feature of being
non-linear and non-local, as it depends on the halo overdensity at the
original Lagrangian positions.  Therefore, the statistical halo
distribution in Lagrangian space, which can also be described in terms
of a hierarchy of bias factors (cf. CLMP), plays the role of the
initial condition in generating the Eulerian halo density field.
       
Aim of the present paper is to provide an exact treatment of the halo
correlation properties in the Lagrangian world, which extends and puts
on sounder bases the results obtained by Mo \& White (1996), as already
sketched in CLMP.  Our approach here is entirely based on the excursion
set version of the PS theory (Peacock \& Heavens 1990; Cole 1991; Bond
\etal 1991, hereafter BCEK), which allows to define halo populations
free of the cloud-in-cloud problem on all relevant scales.

Our formalism has to be considered in every respect as the natural
extension of the BCEK model to include the spatial clustering
properties of the haloes.  Once a cosmological model and a
power-spectrum of density fluctuations are selected the present
formalism allows to construct halo correlations of any order in
Lagrangian space. Here we only give explicit results for the two-point
function, but the generalization to higher order would be
straightforward.

An obvious and important application of this type of study aims at
modeling galaxy clustering at different redshifts.  The key idea being
that the spatial distribution of DM haloes can be a clue for
understanding the clustering properties of luminous objects like
galaxies with different physical properties and at various epochs
(e.g. Kauffmann, Nusser \& Steinmetz 1997; Matarrese \etal 1997;
Moscardini \etal 1997).  The actual relation between the parent DM
haloes and the galaxies, however, depends on the way haloes accrete
mass and on how they grow by the process of merging with the
surrounding haloes. These processes are usually studied in terms of the
so-called `halo merger trees', obtained either by Monte Carlo methods
(e.g. Cole 1991; Lacey \& Cole 1993; Kauffmann \& White 1993; Tozzi,
Governato \& Cavaliere 1996; Somerville \& Kolatt 1997;
Salvador-Sol\'e, Solanes \& Manrique 1997 and references therein) or by
directly looking at the dynamics of particles in cosmological N-body
simulations (e.g. Kauffmann \etal 1997; Roukema \etal 1997; Governato
\etal 1997).  Realizations of halo merger trees can be in fact obtained
by Monte Carlo techniques in the frame of the excursion set theory,
where halo accretion histories are associated to random walks.  The key
idea is to average the linear density field over spheres of decreasing
size centred on each Lagrangian point, thus defining a set of
trajectories of the smoothed field as a function of the resolution
scale. At each cosmic epoch the process of halo formation is here
modeled as the upcrossing of a threshold by these random trajectories.
Therefore, every trajectory gives a detailed description of the mass
accretion history of the corresponding halo.  A merger tree is then
obtained by suitably connecting the accretion histories of all the
progenitors of a given halo (which, in the tree-jargon, are usually
called `branches').

\begin{figure}
%
\centering{
\vbox{
\psfig{figure=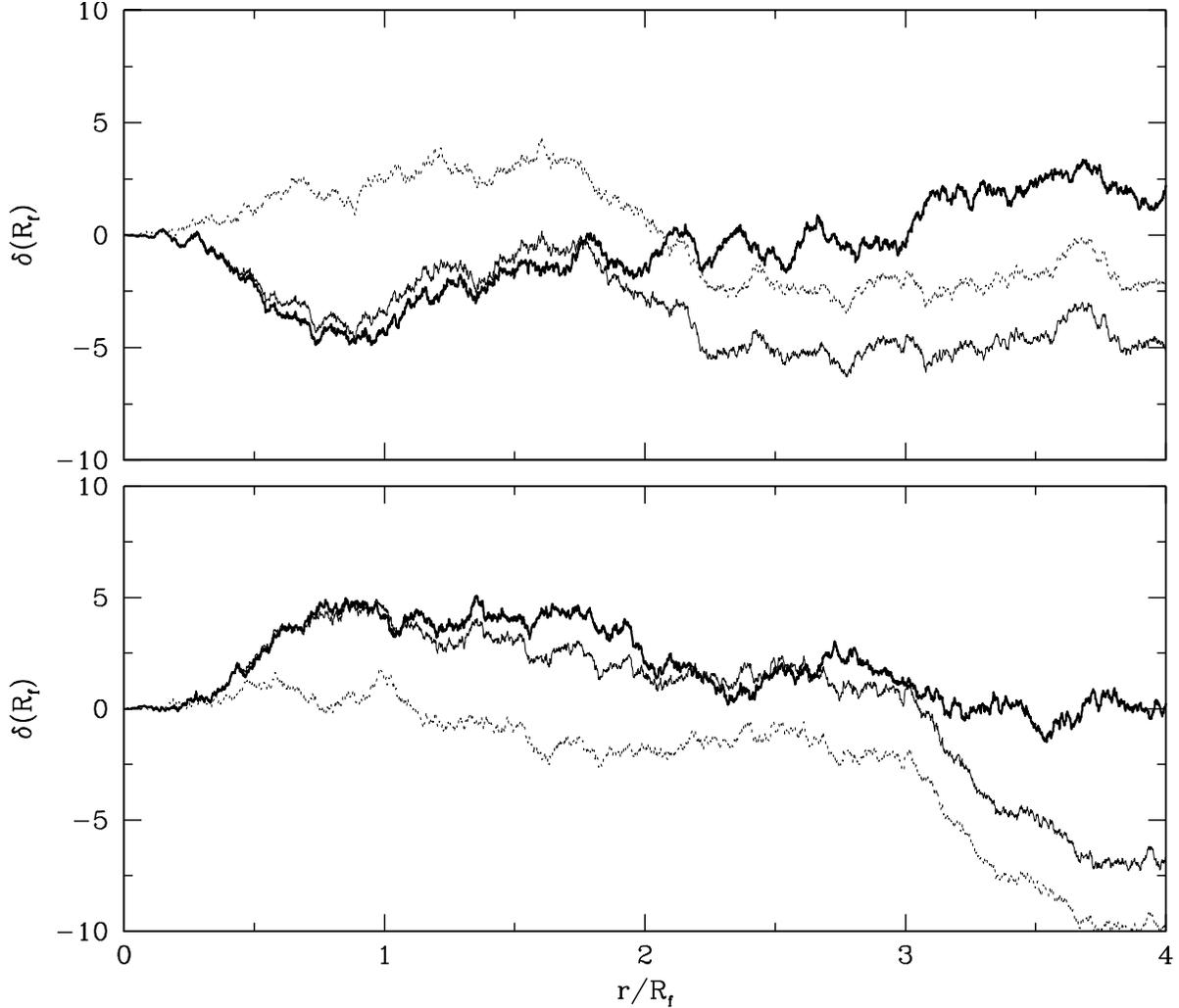,bbllx=21pt,bblly=472pt,bburx=568pt,bbury=693pt,clip=t,width=16cm}
\psfig{figure=fig1.ps,bbllx=21pt,bblly=165pt,bburx=568pt,bbury=422pt,clip=t,width=16cm}
}
}
\caption{Examples of trajectories obtained by smoothing the linear
density fluctuation field with a series of sharp $k$-space filters
(with decreasing resolution scales $R_f=1/k_f$) at two points separated
by the distance $r$. Here we consider $r=4 \,{\rm Mpc}$ and a CDM power
spectrum (with density parameter $\Omega=1$ and present Hubble constant
$H_\0= 50 \vel\,{\rm Mpc}^{-1}$) linearly extrapolated to $\sigma_8=1$.
The {\it heavy continuous} line represents the trajectory associated to
a point $\bfx$ in Lagrangian space.  The trajectory associated to the
point $\bfy$ such that $ |\bfy - \bfx| = r$ is plotted with a {\it
light continuous} line.  The {\it dotted} line is obtained by: {\em i})
considering the trajectory associated to $\bfy$, {\em ii}) artificially
removing any correlation with the trajectory at $\bfx$ and {\em iii})
suitably rescaling the result.  In practice, the dotted line represents
a trajectory which is completely independent of the one associated to
$\bfx$, but which has the same statistical one-point properties.  This
is what is generally used in building up merger trees of dark matter
haloes.  However, the actual trajectories associated to neighbouring
points are strongly correlated when the smoothing length $R_f$ is
comparable to the physical separation between the points $r$.  The same
trajectories tend to become less and less correlated as $R_f$
decreases. It is interesting to note that a memory of the strong
correlation existing for $r/R_f \ll 1$ remains even when $r/R_f \gg
1$. In fact, when $R_f$ is small compared to $r$, the real trajectory
associated to $\bfy$ can be accurately approximated by adding a
constant value to the independent trajectory so that it matches the
random walk at $\bfx$ when $R_f \magcir\,r$.  In other words, since the
trajectories are continuous functions of $R_f$, the value assumed at
$R_f \sim r$ (i.e. at the end of the strongly correlated regime)
represents the initial condition for the random walk at small $R_f$
(i.e. in the independent regime).  This derives from a sort of natural
peak-background split: by evaluating the density fluctuation field in
two points separated by $r$ one samples the same background value
(obtained by summing up all the Fourier components corresponding to
wavelenghts larger than $r$) but different high frequency modes.}
\label{trajectories}
\end{figure}

In the latter approach, however, spatial correlations between different
branches of the tree as well as correlations between different trees
are completely ignored.  The existence of this shortcoming has been
also noticed by Rodrigues \& Thomas (1996) in connection with the
so-called `block model' (Cole \& Kaiser 1988; Cole 1991), which can be
considered as a rudimentary form of merging tree, where haloes can only
grow by doubling their mass. Yano, Nagashima \& Gouda (1996) and
Nagashima \& Gouda (1997) analysed the possible changes induced by
spatial correlations on the PS mass function.
  
The problem within the standard Monte Carlo approach is that each
trajectory is treated as being independent of all the others. In a
general Gaussian field, however, this situation is not realized and,
for power-spectra of cosmological interest, the density fluctuations
turn out to be strongly correlated on small scales. This statistical
property can be clearly detected by looking at a pair of trajectories
associated to two neighbouring points.  In Fig. \ref{trajectories} we
show two trajectories obtained by smoothing a correlated field with a
sharp $k$-space filter.  When the smoothing length is much larger than
the separation between the points in which the trajectories are
computed the density contrast assumes values that are practically
coincident.  On the other hand, for filtering radii much smaller than
the lag that separates the points, the trajectories become indeed
uncorrelated.  The transition from the completely correlated situation
to the totally uncorrelated one is continuous, i.e. the trajectories
become less and less correlated as the smoothing length decreases.  For
comparison, in Fig. \ref{trajectories} we also show the same
trajectories modified by artificially removing any correlation between
them.  It is clear that the joint distribution of first upcrossings at
different points is heavily influenced by the correlations of the
underlying density fluctuation field.

In this paper we will {\em i)} compute the joint two-point distribution
of the first upcrossing events and use the solution to obtain an
approximate analytical expression for the halo-halo correlation
function, {\em ii)} numerically solve suitable sets of spatially
correlated Langevin equations to directly compute the halo two-point
function at different lags.  In a forthcoming paper we will consider
the effects of correlations on the merging histories of dark matter
haloes.

The plan of the paper is as follows. In Section 2 we present a
straightforward extension of the excursion set approach and review the
standard application of this method which allows to determine the mean
mass function of DM haloes. In Section 3 we study the Lagrangian
clustering of haloes identified by our scheme and obtain the halo-halo
correlation function both from an approximate analytic treatment and
from a numerical integration of two spatially correlated Langevin
equations. Section 4 contains a discussion of our results and hints at
some applications of our algorithm to generate spatially correlated
halo merger trees.

\section{Excursion set approach}

In this section we review the formulation of the excursion set
approach, mostly following the treatment given by BCEK.

\subsection{Langevin equation}

Let us assume that the linear density fluctuations form a homogeneous
and isotropic Gaussian random field $\delta({\bf x},z)$, with $\bf x$
the Lagrangian comoving position and $z$ the redshift. In linear theory
one can factor out the redshift dependence as $D(z)\de(\bfx)$, where
$D(z)$ is the linear growth factor (e.g. Peebles 1980). With the
normalization $D(z=0)=1$, $\de(\bfx)$ becomes the mass density
fluctuation linearly extrapolated to the present epoch.

The statistical properties of the Gaussian field $\de(\bfx)$ are
completely specified by the two-point function in Fourier space, which
is related to the power-spectrum $P(k)$ by $\lan \fde (\bfk_1)\, \fde
(\bfk_2)\ran = (2\pi)^3 \ded (\bfk_1+\bfk_2) P(k_1)$. The symbol $\ded$
represents the Dirac delta function, and the brackets $\lan \cdot \ran$
denote ensemble averaging. Our Fourier transform convention is
$\fde(\bfk)=\int d\bfx\,\de(\bfx)\,{\rm e}^{i\bfk\cdot\bfx}$.

We want now to study the statistical properties of the density
fluctuation field at some resolution scale $R_f$. This is introduced by
convolving $\de(\bfx)$ by some filter function $W(|\bfx'-\bfx|, R_f)$,
\be
\de(\bfx, R_f)=\int d\bfx'\,W(|\bfx'-\bfx|, R_f)\,\de(\bfx')=
\f{1}{(2 \pi)^3}
\int d\bfk\, \fW(kR_f)\,\fde(\bfk)\,
{\rm e}^{-i\bfk\cdot\bfx}\;,
\label{filtra}\ee
where $\fW$ is the Fourier transform of the filter.  At each point
$\bfx$ the smoothed field represents the weighted average of
$\de(\bfx)$ over a spherical region of characteristic dimension $R_f$
centred in $\bfx$. The detailed properties of $\de(\bfx, R_f)$ clearly
depend upon the specific choice of window function.  The most commonly
used smoothing kernels are the top-hat filter $W_{TH}(|\bfx|,
R_f)=3\,\Theta(R_f-|\bfx|)/4\pi R_f^3$, where $\Theta(x)$ is the
Heaviside step function, and the Gaussian one $W_{G}(x,R_f)=(2\pi
R_f^2)^{-3/2}\exp(-x^2/2R_f^2)$.  Recently, for convenience of
analysis, top-hat filtering has been also applied in momentum space
$\fW_{SKS}(k, R_f)= \Theta (k_f-k)$, where $k_f=1/R_f$ and
$k_f=|\bfk_f|$. This kernel is generally called sharp $k$-space filter.
While it is easy to associate a mass to real space top-hat filtering
$M_{TH}(R_f)=4\pi\rho_b R_f^3/3$, there is always a bit of
arbitrariness in assigning a mass to the other window functions.  The
most common procedure is to multiply the average density by the volume
enclosed by the filter, obtaining $M_{G}(R_f)= (2\pi)^{3/2}\rho_b
R_f^3$ and $M_{SKS}(R_f)=6\pi^2\rho_b k_f^{-3}$ (Lacey \& Cole 1993).
An alternative procedure, originally introduced by Bardeen \etal
(1986), corresponds to the choice $M_{SKS}(R_f)= 4\pi \rho_b
R_{TH}^3/3$, where $R_{TH}$ is chosen by requiring
$\sigma^2_{SKS}(R_f)=\sigma^2_{TH}(R_{TH})$, and similarly for the
Gaussian filter. In this way one obtains good agreement with numerical
simulations of clustering growth (Lacey \& Cole 1994).
 
In hierarchical bottom-up models of structure formation at each epoch
mass fluctuations have typically undergone non-linear collapse up to
some scale $R_f$, and the first objects that form are those of lower
mass. Haloes of larger mass arise by the merging and accretion of
subunits. The mass distribution deriving from this hierarchical halo
formation process has been successfully modeled by Peacock \& Heavens
(1990), Cole (1991), BCEK and Lacey \& Cole (1993).  In order to mimic
the accretion of matter all these models consider a full hierarchy of
decreasing resolution scales $R_f$. The effect of varying $R_f$ can be
obtained by differentiating eq. (\ref{filtra})
\be
\label{lan}
{\partial \de(\bfx,R_f) \over \partial R_f}=
{1 \over (2\pi)^3} \int d\bfk\,\fde(\bfk)\,
{\partial \fW(kR_f) \over \partial R_f}\,
{\rm e}^{-i\bfk\cdot\bfx} \equiv \eta(\bfx, R_f) \;.
\ee
This has the form of a Langevin equation, which shows how an
infinitesimal change of the resolution scale $R_f$ affects the value of
the density fluctuation field $\delta(\bfx, R_f)$ in the given position
$\bfx$ through the action of the stochastic force $\eta(\bfx, R_f)$. In
the limit $R_f \to \infty$ one has $\de(\bfx;R_f)\to 0$, which can be
adopted as initial condition for our first-order stochastic
differential equation.  Thus, by solving it, we can associate to each
point $\bfx$ a trajectory $\delta(\bfx, R_f)$ obtained by varying the
resolution scale $R_f$. Notice that, since eq. (\ref{lan}) is linear,
$\eta(\bfx, R_f)$ is also a zero mean Gaussian random field, uniquely
specified by its auto-correlation function
\be
\langle \eta(\bfx_1,R_{f1})\, \eta(\bfx_2,R_{f2}) \rangle = {1\over
2\pi^2} \int_0^\infty dk_1\, k_1^2\, P(k_1)\, 
{\p \fW(k_1R_{f1})\over \p R_{f1}}\,
{\p \fW(k_1R_{f2}) \over \p R_{f2}}\,
j_0(k_1r)\;,
\label{eq3}
\ee
where $r=|\bfx_1-\bfx_2|$ and $j_0(x)$ is the zeroth order spherical
Bessel function. Trajectories associated to different neighbouring
points will be statistically influenced by the correlation properties
of the force $\eta(\bfx, R_f)$, i.e. of the underlying Gaussian field
$\de(\bfx)$. On the other hand the coherence of each trajectory along
the $R_f$ direction depends exclusively on the analytic form of the
filter function and vanishes for the sharp $k$-space window
(BCEK). With such a filter, by decreasing the smoothing length one adds
up a new set of Fourier modes of the unsmoothed distribution to
$\delta(\bfx, R_f)$. For a Gaussian field this is completely
independent of the previous increments, and each trajectory
$\delta(\bfx, R_f)$ becomes a Brownian random walk.

In the case of sharp $k$-space filtering the notation greatly
simplifies if we use as time variable the variance of the filtered
density field, $\Lambda \equiv\sigma^2(k_f) = \lan \de(k_f)^2\ran=
(2\pi^2)^{-1} \int_0^{k_f}dk\, k^2\, P(k)$.  In such a case the
stochastic process reduces to a Wiener one, namely
\be \f{\p \de(\bfx, \Lambda)}{\p\Lambda} = \zeta(\bfx, \Lambda) \;,
\label{eq4}
\ee
with $\langle \zeta(\bfx, \Lambda) \rangle=0$ and
\be
\langle \zeta(\bfx, \Lambda_1) \,\zeta(\bfx, \Lambda_2) \rangle = \ded
(\Lambda_1-\Lambda_2)
\label{eq5}
\ee
[see eq. (17) for the spatial correlation]. In the following we will
adopt $\Lambda$ as time variable, unless explicitly stated.  The
solution of the Langevin equation (\ref{eq4}) in an arbitrary point of
space (the position $\bfx$ is here understood), with the initial condition
$\de(\Lambda=0)=0$, is simply $\de(\Lambda)=\int_0^\Lambda
d\Lambda^\prime \zeta(\Lambda ^\prime)$.  By ensemble averaging this
expression one obtains $\langle \de(\Lambda) \rangle= 0$ and $\langle
\de(\Lambda_1) \de(\Lambda_2) \rangle = {\rm min} (\Lambda_1,
\Lambda_2)$, which uniquely determine the Gaussian distribution of
$\de(\Lambda)$. More details can be found in Porciani \etal (1996).

\subsection{Press-Schechter mass function}

Press \& Schechter (1974) proposed a simple model to compute the
comoving number density of collapsed haloes directly from the
statistical properties of the linear Gaussian density field (see also
Doroshkevich 1967). According to the PS theory a patch of fluid is part
of a collapsed region of scale larger than $M(R_f)$ if the value of the
smoothed linear density contrast on the same scale exceeds a threshold
$t_f$. The idea is to use a global threshold in order to mimic
non-linear dynamical effects ending up to halo collapse and
virialization.  An exact value for $t_f$ can be obtained by describing
the evolution of density perturbations according to the spherical
top-hat model.  In this case a fluctuation of amplitude $\de$ will
collapse at a redshift $z_f$ such that $\de(\bfx) =
t_f\equiv\de_c/D(z_f)$. In the Einstein de Sitter universe and during
the matter dominated era the critical value $\de_c$ does not depend on
any cosmological parameter and is given by $\de_c = 3(12 \pi)^{2/3}/20
\simeq 1.686$, whereas, for general cosmologies, it shows a weak
dependence on the value of the density parameter, the cosmological
constant, the Hubble constant, thus on redshift (e.g. Lacey \& Cole
1993).

The PS formula for the mass function has been thoroughly compared with
the outcomes of N-body simulations, finding general good agreement
(e.g. Efstathiou \etal 1988; Gelb \& Bertschinger 1994; Lacey \& Cole
1994).  As mentioned in the Introduction, the model is intrinsically
flawed by the cloud-in-cloud problem, namely the fact that a
fluctuation on a given scale can contain substructures of smaller
scales and the same fluid elements can be assigned, according to the PS
collapse criterion, to haloes of different mass.  Moreover, in a
hierarchical scenario, one expects to find all the mass collapsed in
objects of some scale, while the PS model can account only for half of
it: this problem is intimately related to the fact that in a Gaussian
field only half volume is overdense.  Press and Schechter faced the
problem by simply multiplying their result by a fudge factor of 2.  In
this section we review how the Langevin equations introduced above can
be used to extend the PS theory in such a way to solve both
problems. In the next sections we will use the same approach to compute
the halo-halo correlation function.

The solution of the cloud-in-cloud problem has been given by Peacock \&
Heavens (1990), Cole (1991) and BCEK. Their approach consists in
considering at any given point the trajectory $\delta(R_f)$ as a
function of the filtering radius, and then determining the {\it
largest} $R_f$ at which $\delta(R_f)$ upcrosses the threshold
$t_f(z_f)$ corresponding to the formation redshift $z_f$.  This
determines the largest mass collapsed at that point, all sub-structures
having been erased. So, the computation of the mass function is
equivalent to calculating the fraction of trajectories that first
upcross the threshold $t_f$ as the scale $M$ decreases. The solution of
the problem is enormously simplified for Brownian trajectories, that is
for sharp $k$-space filtered density fields. In such a case one only
has to solve the Fokker-Planck equation for the probability density
$\calW(\de, \Lam)\,d\de$ that the stochastic process at $\Lam$ assumes
a value in the interval $\de, \de+d\de$,
\be
\f{\p \calW(\de, \Lam)}{\p\Lam} = 
\f{1}{2}\,\f{\p^2\calW(\de,\Lam)}{\p\de^2}\; ,
\label{fp}
\ee
with the absorbing boundary condition $\calW(t_f, \Lam)=0$ and initial
condition $\calW(\de, 0)=\de_D(\de)$. The solution is well known
(Chandrasekhar 1943)
\be
\calW(\de, \Lam; t_f)\,d\de =
{1\over \sqrt{2\pi\Lambda}} \left[ \exp \left(-{\delta ^2 \over 2\Lambda}
\right) - \exp \left(-{(\delta -2t_f)^2 \over 2\Lambda} \right)
\right]\,d\de \;.
\label{chandra}
\ee
Defining $\,S(\Lambda, t_f)= \int _{-\infty}^{t_f} \!d\delta\,{\cal
W}(\delta, \Lambda, t_f)\,$ as survival probability of the
trajectories, one obtains the density probability distribution of
first-crossing variances by differentiation
\be
{\cal P}_1 (\Lambda) =-{\p S(\Lam, t_f) \over \p \Lam} =
-{\p  \over \p \Lam}\int_{-\infty}^{t_f}\!d\de\, 
\calW(\de,\Lam; t_f) =
\left[- {1\over 2} {\partial \calW(\delta,\Lambda, t_f)
\over \partial \delta}
\right] _{-\infty}^{t_f}=
{t_f \over \sqrt {2\pi \Lambda^3} }
\exp \left( -{t_f^2 \over 2 \Lambda} \right)\;.
\label{Smir}
\ee
The function ${\cal P}_1(\Lambda) \,d\Lambda$ yields the probability
that a realization of the random walk is absorbed by the barrier in the
interval $(\Lambda, \Lambda+d\Lambda)$ or, by the ergodic theorem, the
probability that a fluid element belongs to a structure with mass in
the range [$M(\Lambda+d\Lambda), M(\Lambda)$]. Finally, the comoving
number density of haloes with mass in the interval $[M, M+dM]$
collapsed at redshift $z_f$ is
\be
\label{genps}
n(M,z_f)\, dM = {\rho_b \over M}\, {\cal P}_1(\Lambda) \,
\left| {d\Lambda \over dM} \right|\, dM \;.
\ee
Inserting the expression of ${\cal P}_1(\Lambda)$ of eq. (\ref{Smir})
in the latter equation one obtains the well-known PS expression for the
mass function
\be
\label{ps}
n(M,z_f)\, d M =
{\rho_b \,t_f(z_f) \over \sqrt {2\pi}}\,
{1\over M^2 \sqrt{\Lam(M)} } \,
\left| {d\ln \Lambda \over d\ln M} \right|\;
\exp \left( - {t_f(z_f)^2 \over 2 \,\Lambda (M)} \right) \,dM\;.
\ee
The original fudge factor of 2 of the PS approach is now naturally
justified.  A generalization of this formalism to simple cases of
non-Gaussian initial conditions has been given by Porciani \etal
(1996).

Previous investigations (e.g. Peacock \& Heavens 1990; BCEK) have shown
that only for sharp $k$-space filtering it is possible to write an
analytic formula for the mass function obtained from the excursion set
approach.  Numerical solutions of the cloud-in-cloud problem with
physically more acceptable smoothing kernels like Gaussian and top-hat
result in mass functions that are a factor of two lower in the
high-mass tail and have different small-mass slopes compared with the
PS formula.  The standard interpretation of this result is that the
excursion set method is not reliable for $M\ll M_\ast$, where $M_\ast$,
defined by $\Lambda(M_\ast)=t_f^2$, is the typical mass collapsing at
$z_f$.

\section{Lagrangian clustering}

In this section we show how the excursion set approach can be extended
to derive the clustering properties of dark matter haloes.
Specifically, we study the evolution of a set of density fluctuation
processes at different spatial locations, as the smoothing scale
progressively shrinks. The trajectories turn out to be intimately
correlated and the joint distribution of the first upcrossing filtering
radii is used to extract the Lagrangian halo-halo correlation function.

\subsection{Two-point correlation function from joint 
upcrossing distribution}

Let us select two points in Lagrangian space $\bfx_1$ e ${\bf x}_2={\bf
x}_1+{\bf r}$. We want to study the evolution of the stochastic
processes $\delta_1 (\Lambda)=\delta ({\bf x}_1,\Lambda)$ and $\delta_2
(\Lambda)=\delta ({\bf x}_2,\Lambda)$ as $\Lambda$ varies. In
particular, let us suppose that we know the joint probability
distribution $\calP_2(\Lambda_1,\Lambda_2;{\bf x}_1, {\bf r}) $ of
those pairs of variances $(\Lambda_1,\Lambda_2)$ that correspond to the
first upcrossing scales of the threshold $t_f$ by the two processes
$\delta_1 (\Lambda)$ and $\delta_2 (\Lambda)$.  Because of the
underlying homogeneity and isotropy, the probability density $\calP_2$
cannot depend on the vector ${\bf x}_1$ and on the orientation of
$\bfr$, i.e.  $\calP_2(\Lambda_1,\Lambda_2; {\bf x}_1, {\bf r}) =
\calP_2(\Lambda_1,\Lambda_2;r)$.  Moreover, by the ergodic theorem one
can identify $\calP_2(\Lambda_1,\Lambda_2;r)$ with the probability
distribution of the pairs $(\Lambda_1,\Lambda_2)$, obtained by randomly
selecting points spatially separated by the distance $r$, within a
given realization of the density field.  Finally, following the
arguments given in the previous section, we can interpret
$\calP_2(\Lambda_1,\Lambda_2;r)\,d\Lambda_1 d\Lambda_2$ as the
probability of finding two points separated by $r$ within two haloes
with mass in the intervals $(M_1-dM_1,M_1)$ and $(M_2-dM_2,M_2)$, as
fixed by the corresponding variance ranges
$(\Lambda_1,\Lambda_1+d\Lambda_1)$ and
$(\Lambda_2,\Lambda_2+d\Lambda_2)$.  As we will discuss in detail in
the next sections, the probability density
$\calP_2(\Lambda_1,\Lambda_2;r)$ can be obtained by integrating the
system of correlated Langevin equations that describe the evolution of
the processes $\delta_1 (\Lambda)$ and $\delta_2 (\Lambda)$.

In order to compute the halo-halo correlation function we adopt the
following procedure. A class of objects is selected by the mass
interval corresponding to the $\Lambda$-range $I \equiv [\Lambda_{\rm
min},\Lambda_{\rm max}]$. The probability of determining two points
separated by $r$ contained within collapsed objects of class $I$ is
\be
\calP_{II}(r)=\int_I\int_I d\Lambda_1\, d\Lambda_2\,
\calP_2(\Lambda_1,\Lambda_2;r)\;.
\ee
Similarly, the probability of finding a point contained in an object of
type $I$ is $ \calP_I=\int_I d\Lam\,\calP_1(\Lam) $. From the
definition of correlation function we then obtain
\be
\xi_{II}^{pts}(r)={\calP_{II}\over \calP_I^2}-1=
{\int _I \int _I   d\Lam_1\, d\Lam_2 \,\calP_2(\Lam_1,\Lam_2;r) \over 
\left[ \int_I d\Lam \,\calP_1(\Lam)\right] ^2}-1\;.
\label{xipts}
\ee
In a similar fashion, considering disjoint classes $I_1$ and $I_2$,
\be
\xi_{I_1I_2}^{pts}(r)={\int _{I_1} \int _{I_2}d\Lam_1\, d\Lam_2\,
\calP_2(\Lam_1,\Lam_2; r) \over  
\int _{I_1}d\Lam \,\calP_1(\Lam)\,\int_{I_2}d\Lam\, \calP_1(\Lam)}-1\;.
\label{xiptsdis}
\ee
We stress that the quantities $\xi_{II}^{pts}(r)$ and
$\xi_{I_1I_2}^{pts}(r)$ are the Lagrangian correlations of the mass
elements contained in the collapsed haloes.  This quantity has been
recently used by Bagla (1997) to study the evolution of galaxy
clustering.  However, we are ultimately concerned with the calculation
of the halo correlations $\xi_{II}^{hh}(r)$, so we have to properly
weigh the statistical contribution for each extended halo. The problem
shows particularly simple if we adopt the PS theory. Indeed, in this
case, the sets of points where the first upcrossings occur at the same
$\Lam$ are point-like disconnected regions (see Bond \& Myers 1996 for
a different approach). Only in a statistical sense they originate
collapsed haloes, each contributing by $1/V(\Lam)$, where
$V(\Lam)\equiv M/ \rho_b $ is the typical Lagrangian volume of an
object of mass $M$ associated to the variance $\Lam$. Therefore,
adopting the same procedure used to calculate the mass function, the
mean number density of collapsed objects of scale $\Lam$ becomes
$n(\Lam)=\calP_1(\Lam)/ V(\Lam)$.  Similarly, for the distribution of
pairs at distance $r$, we define
\be
 n_{2}(\Lam_1,\Lam_2;r)={\calP_2(\Lam_1,\Lam_2;r) \over V(\Lam_1)
 V(\Lam_2)} \;.
\ee
The idea is, once again, to allow for the finite size of the haloes.  The
halo-halo correlations become, respectively,
\be
\xi_{II}^{hh}(r)={n_{II}\over n_I^2}-1\equiv{\int _I \int _I 
d\Lam_1 d\Lam_2 \,
n_2(\Lam_1,\Lam_2;r) \over
 \left[ \int _I d\Lam\, n(\Lam)\right]
^2}-1\; ,
\label{xi}
\ee
and
\be
\xi_{I_1I_2}^{hh}(r)={\int _{I_1} \int _{I_2}d\Lam_1 d\Lam_2\,
n_2(\Lam_1,\Lam_2,r)\over  
\int _{I_1}d\Lam\, n(\Lam) \int_{I_2}d\Lam\, n(\Lam)}-1\;.
\label{ximista}
\ee

\subsection{Correlated Langevin equations: sharp k-space filtering}

In Section 3.1 we showed that, in order to obtain the halo correlation
function, it is crucial to know the joint distribution
$\calP_2(\Lam_1,\Lam_2;r)$. This quantity can be obtained by solving
the system of equations governing the evolution of the pair of
correlated processes $\delta_1(\Lam)$ and $\delta_2(\Lam)$:
\be
\left\{
\begin{array}{l}
{\displaystyle {\p \delta_1(\Lam) \over  \p\Lam}=\zeta_1(\Lam)\;,
\;\;\;\;\;\;\;\;\;\;\;\;\;\;\;\;\;   
\delta_1(0)=0 }\;, \\ \\
{\displaystyle {\p \delta_2(\Lam) \over  \p\Lam}=\zeta_2(\Lam)\;,
\;\;\;\;\;\;\;\;\;\;\;\;\;\;\;\;\;
\delta_2(0)=0} \;, \\ \\
\langle \zeta _1(\Lam) \rangle=\langle \zeta _2(\Lam) \rangle=0\;,
\;\;\;\;\;\;\;\;
\zeta_1\;\,{\rm and}\;\, \zeta_2\;\,
{\rm Gaussian\;\;\, processes}\;, \\ \\
{\displaystyle  \langle \zeta_1(\Lam)\, \zeta_1(\Lam')\rangle =
\langle \zeta_2(\Lam)\, \zeta_2(\Lam')\rangle =
 \delta_D(\Lam-\Lam')}\;,\\ \\
{\displaystyle \langle \zeta_1(\Lam) \,\zeta _2(\Lam') \rangle =
\f{\p\xi(r;\Lam)}{\p\Lam}\,\delta_D(\Lam-\Lam')}\;.
\end{array}
\right.
\label{lange2xi}
\ee
The latter equation, obtained by introducing sharp $k$-space filtering
in equation (\ref{eq3}), completes the definition of the stochastic
field $\zeta({\bf x},\Lam)$ given in equations (\ref{eq4}) and
(\ref{eq5}).  Here $\xi(r;\Lam)$ is the linear two-point correlation
function for the mass density fluctuations smoothed on the scale
$R_f\equiv 1/k_f$ associated to the variance $\Lam$
\be
\xi(r; \Lam)\equiv{1\over 2 \pi ^2}\int_0^{k_f\!(\Lam)}dk\,k^2
\,P(k)\;j_0(kr)\;
\ee
and one has
\be 
\frac{\p\xi(r;\Lam)}{\p\Lam} =j_0[k_f(\Lam)\,r]\;.
\ee
By integrating the above differential equations and averaging over the
ensemble one obtains the unconstrained probability distribution
\be
\displaystyle \calW_r(\delta_1,\delta_2; \Lam)=
{1\over 2 \pi \sqrt{\Lam^2-\xi(r;\Lam)^2}}\,
\exp {\left[-{\Lam(\delta_1^2+\delta_2^2)-2\,\xi(r;\Lam)\,
\delta_1 \delta_2
\over 2 \left[ \Lam^2-\xi(r;\Lam)^2 \right]}\right] }\;. 
\label{gaus2} 
\ee
This function solves the two-dimensional Fokker-Planck equation
associated to the Langevin equation (\ref{lange2xi}), namely
\be
{\p \calW_r(\de_1,\de_2; \Lam) \over \p \Lam}=
\f{1}{2}\left[
\f{\p^2}{\p\de_1^2}+
\f{\p^2}{\p\de_2^2}
\,+\,
2\,\f{\p \xi(r; \Lam)}{\p \Lam}\,\f{\p^2}{\p\de_1\p\de_2}\right]
\,\calW_r(\delta_1,\delta_2; \Lam) \;, 
\label{p2}
\ee
with initial condition 
$\calW_r(\de_1,\de_2,\Lam=0)=\ded(\de_1)\,\ded(\de_2)$. The problem of
finding the distribution of the {\it first} upcrossings of the
threshold $t_f$ by the binary process $\{\de_1, \de_2\}$ reduces to
that of imposing proper boundary conditions to the equation (\ref{p2}).
As done for the one-dimensional Fokker-Planck equation, we adopt the
absorbing barrier approach.  However, in the two-dimensional case we
are considering, the distribution $\calP_2(\Lam_1,\Lam_2;r)$ cannot be
eventually obtained from $\calW_r(\de_1,\de_2; \Lam)$ simply by
differentiation. This is because the whole binary system automatically
disappears as soon as when one `Brownian particle' is first absorbed.
Nonetheless, the joint distribution $\calP_2(\Lam_1,\Lam_2;r)$ can be
in principle calculated by a two-step procedure as follows.

Assuming the same initial condition, one solves the Fokker-Planck
equation with absorbing barriers at $\de_1=t_f$ and $\de_2=t_f$, thus
finding the survival probability density for the pairs which have never
crossed the thresholds. Having found this quantity one can compute the
probability current through each point,
\be
\bfJ(\de_1,\de_2; \Lam)=-\f{1}{2}\left(
\f{\p\calW_r}{\p \de_1}
+\f{\p\xi(r; \Lam)}{\p \Lam}\,\f{\p\calW_r}{\p \de_2}\,,\;\;\;
\f{\p\xi(r; \Lam)}{\p \Lam}\,\f{\p\calW_r}{\p \de_1}
+\f{\p\calW_r}{\p \de_2}
\right)\;.
\ee
On a boundary wall, e.g. $\de_1=t_f$, where $\calW_r(t_f,\de_2;
\Lam)=0$ (implying $\p\calW_r/\p \de_2=0$), this reduces to
\be
\bfJ(t_f, \de_2; \Lam)=-\f{1}{2}\left(
\left.\f{\p\calW_r}{\p \de_1}\right|_{t_f}\,,\;\;\;
\f{\p\xi(r;\Lam)}{\p \Lam}\,\left.\f{\p\calW_r}{\p \de_1}\right|_{t_f}
\right)\;.
\ee
The flux through any point of the barrier $\de_1=t_f$ is then given by
the scalar product $\bfJ \cdot {\bf n}$, where ${\bf n}\equiv (1,0)$ is
the unit vector perpendicular to the absorbing wall,
\be
\calF_r(t_f, \de_2; \Lam)= -\f{1}{2}\,
\left.\f{\p\calW_r}{\p \de_1}\right|_{t_f}\;.
\label{flux}
\ee
The quantity $\calF_r(t_f, \de_2; \Lam)\, d\de_2$ represents the
probability that the pair of processes ($\de_1$,$\de_2$) leave the
permitted region passing through the `gate' $[(t_f,\de_2),
(t_f,\de_2+d\de_2)]$ at the time $\Lam$.  This flux contains all the
information we need for the computation of
$\calP_2(\Lam_1,\Lam_2;r)$. In fact, once $\de_1$ has crossed the
barrier at $\Lam_1$, for $\Lam>\Lam_1$ we are interested in studying
only the evolution of the surviving process $\de_2$ up to its first
upcrossing through the boundary $\de_2=t_f$.  Therefore, since we are
considering Brownian trajectories, free of correlations along the
$\Lam$ axis, for $\Lam>\Lam_1$ the evolution of $\de_2$ is simply
governed by its own Langevin equation, and its probability distribution
can be calculated from the one-dimensional Fokker-Planck equation
(\ref{fp}), with absorbing boundary $\de_2=t_f$, {\it but} with initial
condition (at $\Lam=\Lam_1$) $\de_{2\ast}\equiv\de_2(\Lam_1|
\de_1=t_f)$.  Thus, by simply modifying eq. (\ref{Smir}), we find that
the distribution of the variances $\Lam_2$ associated to first
upcrossing events of the threshold by the process $\de_2$, given that
$\de_1$ upcrossed the critical level at $\Lam_1$, is:
\be
{\cal P}_1(\Lam_2-\Lam_1,t_f-\de_{2\ast})=
{(t_f-\de_{2\ast}) \over \sqrt {2\pi}~ (\Lam_2-\Lam_1)^{3/2} }
\exp \left[ -{(t_f-\de_{2\ast})^2 \over 2 (\Lam_2-\Lam_1)} \right]\;.
\label{condl1l2}
\ee
The joint distribution $\calP_2(\Lam_1,\Lam_2; r)$ is 
eventually obtained by a convolution
\be
\calP_2(\Lambda_1,\Lambda_2; r)=\int_{-\infty}^{t_f}d\de_2\,
\calF_r(t_f, \de_2; \Lambda_1)\,{\cal P}_1(\Lambda_2-\Lambda_1, t_f-\de_2)+
\int_{-\infty}^{t_f}d\de_1\,
\calF_r(\de_1, t_f; \Lambda_2)\,{\cal P}_1(\Lambda_1-\Lambda_2, t_f-\de_1)\;, 
\label{convo}
\ee
where the first and second integrals on the r.h.s. represent,
respectively, the contributions of those pairs for which
$\Lam_2\geq\Lam_1$ and $\Lam_2<\Lam_1$.

However, this formal expression is useless unless we solve for the
probability density $\calW_r$.  There are two cases in which the
calculation of $\calW_r$ is trivial: one for $r \to \infty$ and the
other for $r \to 0$. The general case of finite non-zero lag $r$ is
far more complex, as we will discuss.

\subsubsection{Perfectly uncorrelated processes}

At infinite lag ($r \to \infty$) the two processes become independent
and the solution of the Fokker-Planck equation is
\be
\calW_\infty(\de_1,\de_2; \Lam)=\calW(\de_1,\Lam; t_f)\,
\calW(\de_2,\Lam; t_f) \;, 
\ee
where $\calW(\de,\Lam; t_f)$ denotes the probability distribution for a
one-dimensional process with absorbing boundary at $t_f$ given in
eq. (\ref{chandra}).  This solution consists of a linear superposition
of four unconstrained independent density distributions deriving from
different initial conditions. In practice one has to consider the
`real' initial distribution $\ded(0,0)$, an image source
$\ded(t_f,t_f)$ and two image sinks $\ded(t_f,0)$ and $\ded(0,t_f)$,
that is to say
\be
\calW_\infty(\de_1,\de_2;\Lam)=\calG_\infty(\delta_1,\delta_2;\Lam)-
\calG_\infty(\de_1-2t_f,\de_2;\Lam)-
\calG_\infty(\de_1,\de_2-2t_f;\Lam)+
\calG_\infty(\de_1-2t_f,\de_2-2t_f;\Lam)\;,
\label{images}
\ee
where $\calG(\de_1,\de_2;\Lam)\equiv[2\pi\Lam]^{-1}
\exp[-(\de_1^2+\de_2^2)/2 \Lam]$ is the solution of the two-dimensional
Fokker-Planck equation with natural boundary conditions: $\lim_{\de_i
\to \infty} \calG(\de_1,\de_2;\Lam)=0, i=1,2$.  Obviously, using
eq. (\ref{convo}) we obtain
\be
\calP _2(\Lam_1,\Lam_2;r\to \infty)=\calP _1(\Lam_1,t_f)
\calP _1(\Lam_2,t_f) \;,
\ee
that, inserted in equations (\ref{xipts}) and (\ref{xiptsdis}) or in
equations (\ref{xi}) and (\ref{ximista}), gives, as expected for
infinite lag, $\xi^{pts}=\xi^{hh}=0$.

\subsubsection{Perfectly correlated processes}

When $r\to 0$ the two processes become more and more correlated so
that, in the limit, we end up with a one-dimensional problem.  In this
case the solution of the Fokker-Planck equation is
\be
\calW_0(\de_1,\de_2;\Lam)=\calW(\de_1+\de_2,4 \Lam;t_f)\,
\ded(\de_1-\de_2)\;,
\ee
where $\calW(\de, \Lam; t_f)$ denotes the probability distribution
given in eq. (\ref{chandra}).  Expanding this expression as a
superposition of Green's functions, by using the method of image
sources, we obtain
\be
\calW_0(\de_1,\de_2; \Lam)=\calG_0(\de_1,\de_2; \Lam)-
\calG_0(\de_1-2t_f,\de_2-2t_f; \Lam) \;,
\ee
where $\calG_0(x,y; \Lam)\equiv\calW(x, \Lam; t_f)\ded(x-y)$.  In this
case, for the joint distribution of first upcrossing variances, we find
\be
\calP_2(\Lam_1,\Lam_2; r\to 0)=\calP_1(\Lam_1,t_f)\,\ded(\Lam_1-\Lam_2)\;.
\ee
Consequently, we obtain $\xi^{pts}_{II}\to
[1/\int_I\,d\Lam\,\calP(\Lam)] -1$ and $\xi^{hh}_{II} \to \{\int_I\,
d\Lam\, P(\Lam)/V(\Lam)^2 / [\int_I \,d\Lam\,P(\Lam)/V(\Lam)]^2 \}-1$.
Similarly, $\xi^{pts}_{I_1 I_2} \to 0$ and $\xi^{hh}_{I_1 I_2}=0$.

\subsubsection{Approximate general solution}

In the limiting cases just discussed we were able to account for the
boundary conditions by writing the solution in terms of a superposition
of image-distributions, each one solving the Fokker-Planck
equation. Regrettably, the position and the sign of the image sources
of probability come out dependent on the correlation between the
processes (i.e. on $r$). This fact suggests that we cannot write a
general analytical solution by simply applying the image method.
However, we will build here a simple function that satisfies the
absorbing boundary conditions being also an accurate approximation for
the solution of the correlated diffusion equation. In the next section
we will test the accuracy of this solution against the numerical one.

It is evident that, for small separation $r \ll R_f$ (we remind the
reader that we change the smoothing radius $R_f$ at fixed separation
$r$), the perfectly correlated solution will represent a very good
approximation to the true one.  This can be easily deduced by the
following argument.  The two Gaussian stochastic processes
$\Sigma(\Lam)=\de_1(\Lam)+ \de_2(\Lam)$ and
$\Delta(\Lam)=\de_1(\Lam)-\de_2(\Lam)$ are statistically independent,
i.e. $\lan \Sigma(\Lam) \Delta(\Lam) \ran=0$.  The variances of their
unconstrained probability distributions are, respectively,
$\sigma^2_{\Sigma}=2\,[\Lam+\xi(r;\Lam)]$ and
$\sigma^2_{\Delta}=2\,[\Lam-\xi(r;\Lam)]$.  Therefore, for $r \ll R_f$
[i.e. for $\Lam \ll \sigma^2(r)$, where $\sigma^2(r)$ denotes the
variance of the mass density fluctuations smoothed on the scale $r$]
where $\xi(r;\Lam) \simeq \Lam$, we have $\sigma^2_{\Delta} \simeq 0$
and the probability distribution of the variable $\Delta$ is
practically a Dirac delta function centred on the zero value. This
corresponds to the perfectly correlated situation. However, this regime
is not interesting for computing the halo correlation function since
the condition $r \ll R_f$ implies that the points in which we follow
the trajectories that upcross the threshold are involved in the
collapse of the same halo.

\begin{figure}
\centering{
\hbox{
\psfig{figure=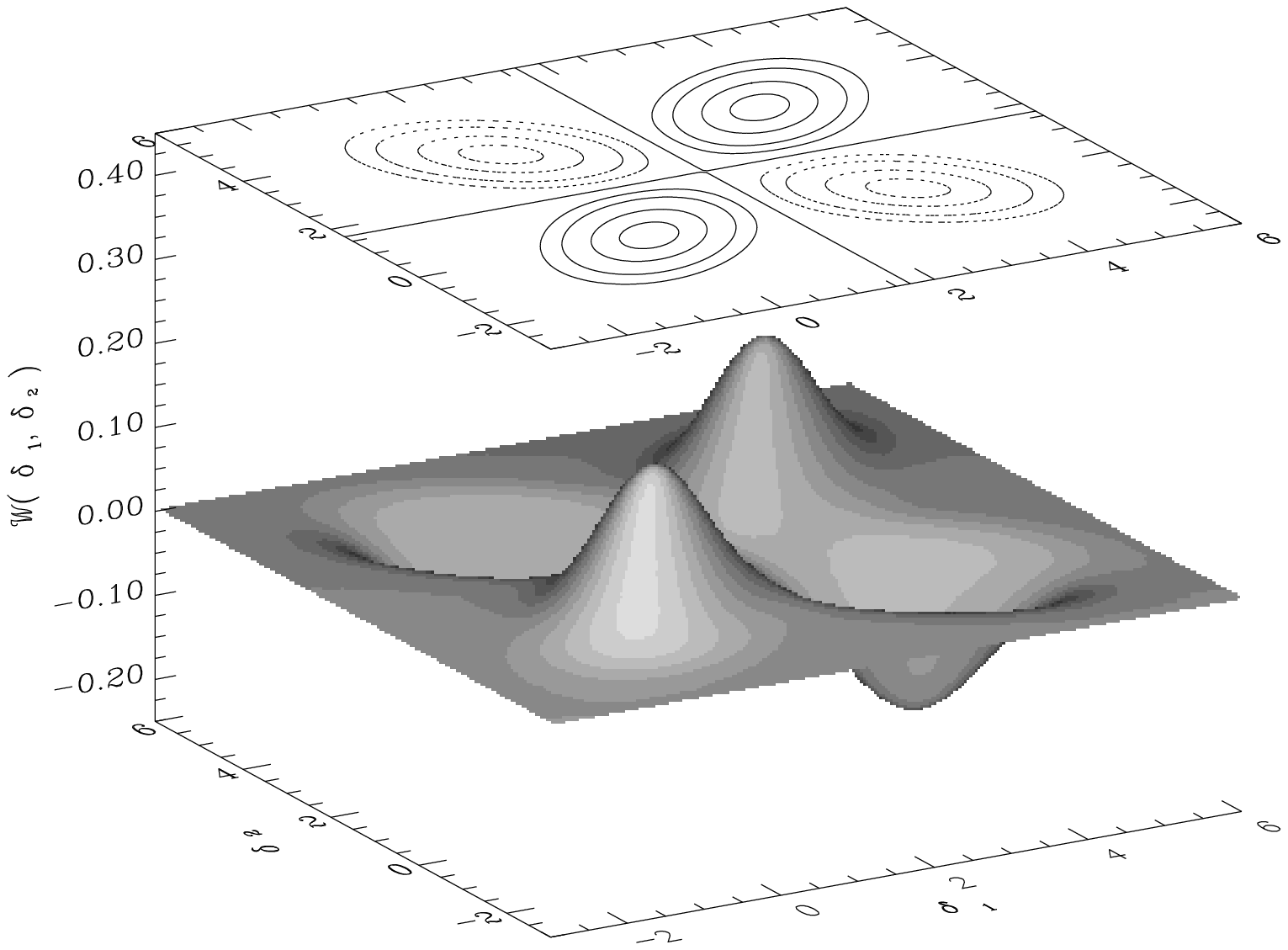,bbllx=21pt,bblly=21pt,bburx=470pt,bbury=352pt,clip=t,height=7cm,width=8cm}
\psfig{figure=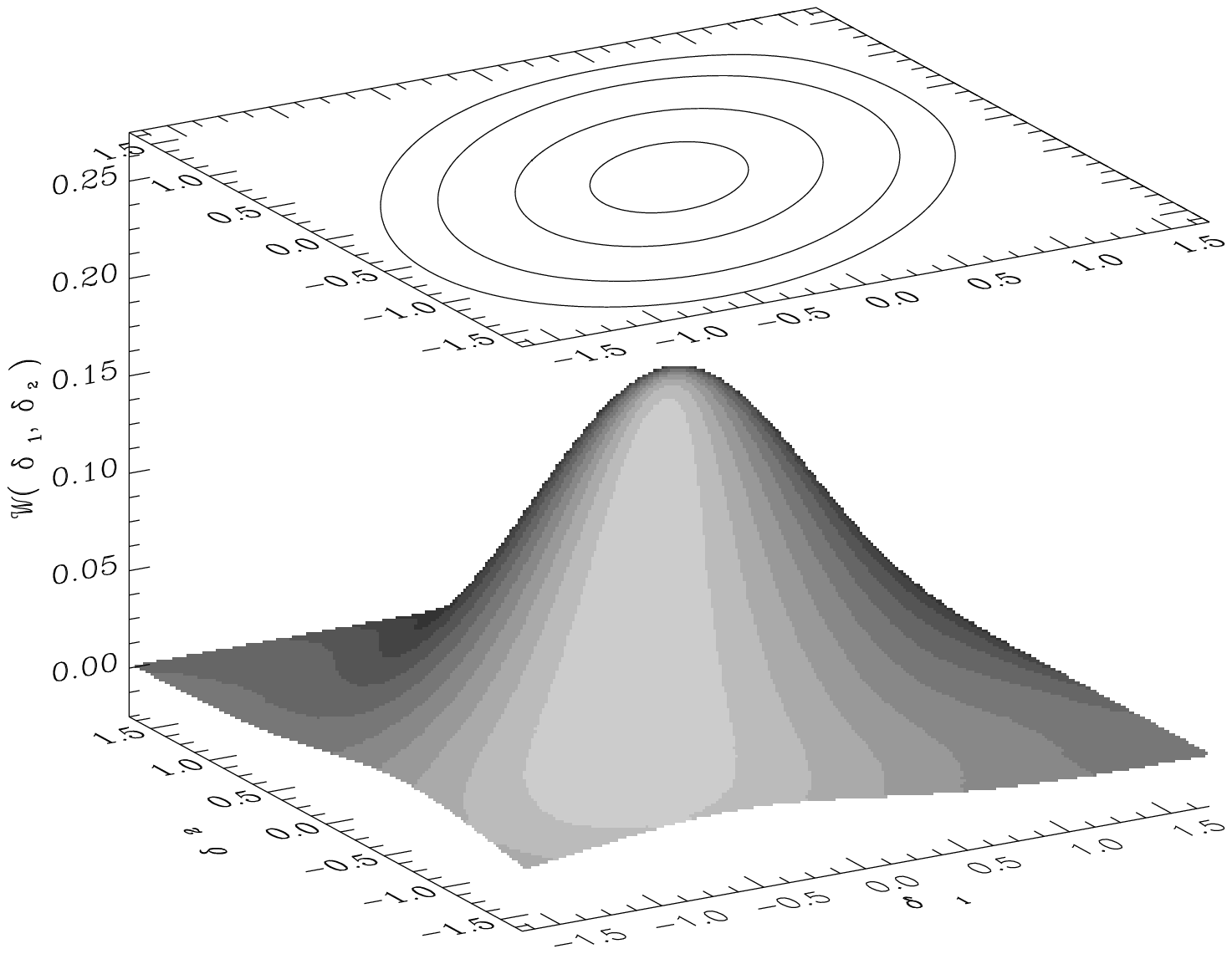,bbllx=32pt,bblly=21pt,bburx=461pt,bbury=352pt,clip=t,height=7cm,width=8cm}
}
}
\caption{The approximate solution of the Fokker-Planck equation given
in eq. (\ref{approxsol}) is plotted in the left panel (here $\Lam=1$
and $\xi=0.4$).  The probability density in the physical region is
blown up in the right panel.  In the contour plots the continuous lines
correspond to the levels 0,~0.025,~0.05,~0.1,~0.15 (starting from
outside), while the dotted lines refer to the same values multiplied by
$-1$.  The figure clearly shows that the proposed solution satisfies
the absorbing boundary conditions for $\de_1=t_f$ and $\de_2=t_f$ (here
$t_f=1.686$).  Moreover, the orientation of the iso-probability curves
in the allowed region (these are practically ellipses with their major
axes lying on the line $\de_1=\de_2$) indicates that the approximate
$\calW(\de_1,\de_2;\Lam)$ has the right correlation properties.}
\label{contour}
\end{figure}

On the other hand, for $r \gg R_f$ (i.e. for $\Lam \gg \sigma^2(r)$),
we can replace in eq. (\ref{p2}) $\xi(r;\Lam)$ with the {\it
unsmoothed} linear mass-correlation $\xi_m(r)$. Therefore, eq.
(\ref{p2}) simply becomes the uncorrelated two-dimensional diffusion
equation that can be easily solved using the image method.  Our
ansatz for the full $\calW_r(\de_1,\de_2;\Lam)$ is then obtained by
keeping the analytic form of the solution just obtained for $r \gg R_f$
but inserting in it the correlation function $\xi(r;\Lam)$ to replace
its large lag limit $\xi_m(r)$. Therefore, we have
\be
\calW_r(\de_1,\de_2; \Lam)=\calG_r^+(\delta_1,\delta_2; \Lam)-
\calG_r^-(\de_1-2t_f,\de_2; \Lam)-\calG_r^-(\de_1,\de_2-2t_f; \Lam)+
\calG_r^+(\de_1-2t_f,\de_2-2t_f;\Lam)\;,
\label{approxsol}
\ee
where 
\be
\calG_r^\pm=
{1\over 2 \pi \sqrt{\Lam^2-\xi(r;\Lam)^2}}\,
\exp {\left[-{\Lam(\delta_1^2+\delta_2^2)-2\,[\pm \xi(r;\Lam)]\,
\delta_1 \delta_2
\over 2 \left[ \Lam^2-\xi(r)^2 \right]}\right] }\;.
\label{approxgreen}
\ee
By using the symbol $\calG_r^\pm$ we want to emphasize the correlation
properties of the adopted Green's function: $\calG_r^+$ is a correct
solution of the diffusion equation, while $\calG_r^-$ does not solve
it. However, to satisfy the boundary conditions, we need to insert it
twice in the solution. The probability distribution given in
eq. (\ref{approxsol}) will be a valid approximation to the proper one
provided the term
\be
\f{\p\xi(r;\Lam)}{\p\Lam} \f{\p^2}{\p\de_1 \p\de_2} 
\left[\calG_r^-(\de_1-2t_f,
\de_2; \Lam)+\calG_r^-(\de_1,\de_2-2t_f; \Lam)\right]
\ee
can be neglected compared to the $\Lambda$-derivative of the expression
in eq. (\ref{approxsol}).  A three-dimensional representation of this
approximate solution and of its contour levels is given in
Fig. \ref{contour}.  Inserting eq. (\ref{approxsol}) into equations
(\ref {flux}) and (\ref{convo}) we obtain
\ba
\calP_2(\Lambda_1,\Lambda_2;r) & = & \f{t_f^2 \Lambda_1 \Lambda_2+\left[
 \Lambda_1 \Lambda_2 -t_f^2(\Lambda_1+\Lambda_2) \right]
\xi(r;\Lam_{\rm m})+t_f^2 \,\xi(r;\Lam_{\rm m})^2-\xi(r;\Lam_{\rm m})^3}
{2 \pi \left[\Lambda_1 \Lambda_2 -\xi(r;\Lam_{\rm m})^2 \right]^{5/2}} 
\times
\nonumber \\
&\times&
\exp{\left[-\f{t_f^2}{2}\f{\Lambda_1+\Lambda_2-2\,\xi(r;\Lam_{\rm m})}
{\Lambda_1 \Lambda_2 -\xi(r;\Lam_{\rm m})^2}\right]} \;,
\label{appp2}
\ea
where $\Lam_{\rm m}={\rm min}(\Lam_1, \Lam_2)$.
By using this expression to compute the halo-halo correlation function
between objects selected in infinitesimal mass ranges, we obtain
\be
\xi^{obj}(r)\equiv \xi^{pts}(r)=\f{\calP_2(\Lambda_1,\Lambda_2;r)}
{\calP_1(\Lambda_1) \calP_1(\Lambda_2)}-1 \;,
\label{appsol}
\ee
whose explicit expression is reported in Appendix A, eq. (A1).  We can
also expand the halo two-point function in powers of the filtered mass
auto-correlation function,
\be
\xi^{obj}(r)\equiv \xi^{pts}(r)=\sum_{n=1}^{\infty}
\f{1}{n!}\,b_n(\Lambda_1)\,b_n(\Lambda_2)\,\xi(r;\Lam_{\rm m})^n \;,
\label{serie}
\ee
where the factors $b_n(\Lambda)$ coincide with the Lagrangian bias
coefficients introduced in eq. (15) of CLMP.  It is important to stress
that eq. (\ref{appsol}), with the ansatz for $\calP_2$ provided by
eq. (\ref{appp2}) and $\calP_1$ given by eq. (\ref{Smir}), represents
an approximation to the exact form of the halo-halo correlation
function which can be obtained by numerically integrating the
correlated Langevin equations, as discussed in the next section.  Other
approximations have been suggested by CLMP following two different
approaches: {\em i}) defining a local halo counting operator acting on
the underlying Gaussian density field [their equation (14), which is
also reported in Appendix A, eq. (A2)], {\em ii}) using the
peak-background split to define the halo overdensity free of the
cloud-in-cloud problem down to the background scale. Both these models
give the same series of Lagrangian bias factors of eq. (\ref{serie})
provided the lag is a few times larger than the Lagrangian halo
size. Let us explicitly write the first four bias factors
\ba
b_1(\Lam)&=&\f{t_f}{\Lam}-\f{1}{t_f} \;,\\ \nonumber
b_2(\Lam)&=&\f{t_f^2}{\Lam^2}-\f{3}{\Lam}\;, \\ \nonumber
b_3(\Lam)&=&\f{t_f^3}{\Lam^3}-\f{6t_f}{\Lam^2}+\f{3}{t_f\Lam}\;, 
\\ \nonumber
b_4(\Lam)&=&\f{t_f^4}{\Lam^4}-\f{10 t_f^2}{\Lam^3}+\f{15}{\Lam^2} \;. 
\ea
The linear bias term, that for $M \ne M_\ast$ dominates the halo
correlation at large separation, coincides with the one obtained by Mo
\& White (1996). This implies that, in this limit, haloes with $M >
M_\ast$ have $b_1>0$, i.e.  are biased with respect to the mass
distribution in Lagrangian space.  Notice that $b_1$ can be very large
when $M\gg M_\ast$.  On the contrary, objects with $M < M_\ast$ have
$-1/t_f<b_1<0$, i.e. are moderately antibiased.  In the limiting case
$M=M_\ast$, $b_1 = 0$ and the leading term of $\xi^{hh}$ is
proportional to $\xi(r;\Lam_{\rm m})^2$, implying much lower halo
correlations compared with different mass ranges.
\footnote{Note, however, that in the Eulerian case the linear bias term
is non-zero also for $M\sim M_\ast$ (e.g. CLMP).}

\subsection {Monte Carlo simulations}

In order to check the validity of the approximate solution introduced
in the previous section we solved numerically for the joint
distribution of first upcrossing variances by integrating our spatially
correlated Langevin equations.  We stress that this method gives the
exact halo-halo correlation function in the excursion set approach,
consistently completing the PS analysis of the mass function.

The stochastic differential equations (\ref{lange2xi}) are equivalent
to the integral equations
\be
\left\{
\begin{array}{l}
{\displaystyle {\de_1(\Lam+\gamma)-\de_1(\Lam)}=
\int_\Lam^{\Lam+\gamma} 
d\Lam'\,
\zeta_1(\Lam')\;,
\;\;\;\;\;\;\;\;\;\;\;\;\;\;\;\;\;   
\delta_1(0)=0 }\;, \\ \\
{\displaystyle {\de_2(\Lam+\gamma)-\de_2(\Lam)}=
\int_\Lam^{\Lam+\gamma} 
d\Lam'\,
\zeta_2(\Lam')\;,
\;\;\;\;\;\;\;\;\;\;\;\;\;\;\;\;\;
\delta_2(0)=0} \;,
\end{array}
\right.
\label{ISE}
\ee
where the statistical properties of the Gaussian processes $\zeta_1$
and $\zeta_2$ are given in eq. (\ref{lange2xi}).  The general procedure
used to solve numerically a stochastic differential equation replaces
the equivalent integral equation by its expansion in power series of
$\sqrt \gamma$, truncates the series after a selected number of terms
and gives a rule for computing each term that is considered.  To
control the effect of the temporal discretization an extrapolation of
the results for $\gamma \to 0$ is often required (Greiner, Strittmatter
\& Honerkamp 1988).  Fortunately, in the case of a set of Wiener
processes this procedure can be greatly simplified. In fact, by
integrating over a finite timestep $\gamma$, equations (\ref{ISE})
simply give
\be
\left\{
\begin{array}{l}
{\displaystyle {\de_1(\Lam+\gamma)-\de_1(\Lam)}=
a_{11}(\gamma,\Lam)\,G_1\;,
\;\;\;\;\;\;\;\;\;\;\;\;\;\;\;\;\;\;\;\;\;\;\;\;
\;\;\;\;\;\;\;\;\;\;\;\;\;\;\;
\delta_1(0)=0 }\;, \\ \\
{\displaystyle {\de_2(\Lam+\gamma)-\de_2(\Lam)}=
a_{21}(\gamma, \Lam) \,G_1 +
a_{22}(\gamma, \Lam) \,G_2\;,
\;\;\;\;\;\;\;\;\;\;\;\;\;\;\;\;\;
\delta_2(0)=0} \;, \\ \\
a_{11}(\gamma, \Lam)^2=a_{21}(\gamma, \Lam)^2+
a_{22}(\gamma, \Lam)^2=
\gamma         \;, \\ \\
a_{11}(\gamma, \Lam)\,a_{21} (\gamma, \Lam)= \xi(r;\Lam+\gamma)-
\xi(r;\Lam)\;,
\end{array}
\right.
\label{DE}
\ee
where the $G_i$ are independent Gaussian variables with zero mean and
unit variance.  This set of equations gives the fundamental recipe to
produce trajectories that are obtained iterating eq. (\ref{DE}) at each
time step by modeling the $G_i$ terms with Gaussian pseudo-random
numbers. To generate these normally distributed deviates in first
passage problems, where rare fluctuations are crucial, it is of fundamental
importance to adopt a method that is accurate even for less probable
events.  For this reason we adopted the Box-Muller algorithm
(e.g. Press \etal 1992), that is rather slow but produces an unbiased
Gaussian distribution (in practice the limited precision of numerical
computation only affects the extreme tail behaviour).  A moderate
speeding up (roughly $20 \% $) is obtained by modifying the algorithm
to use a rejection technique (Knuth 1981; Press \etal 1992).

To estimate the first-passage time distribution one first solves the
discretized stochastic equation starting at the initial point and
terminates the simulation of a trajectory as soon as the boundary is
reached. In order to avoid the resulting distribution being influenced
by the temporal discretization one has to account for possible
intra-step crossings.  In fact, the conditions $\de(\Lam)<t_f$ and
$\de(\Lam+\gamma)<t_f$ do not guarantee that the process $\de$ has ever
crossed the threshold during the time interval $\gamma$.\footnote{This
has a striking analogy with the solution of the cloud-in-cloud problem
given by BCEK.}  For this reason, the simple algorithm of choosing as
first-crossing time that corresponding to the first step at which $\de
> t_f$ is very inaccurate, unless one uses very small time-steps.  This
problem can be solved by performing a small Monte Carlo test at each
time step as shown by Strittmatter (1987). In this way we obtain high
precision even using larger time-steps, therefore reducing the CPU
time.

For a given power-spectrum, once the value for the critical threshold
$t_f$ and the lag $r$ are selected, the algorithm just introduced gives
a pair of first upcrossing variances for each realization of the
processes $\de_1$ and $\de_2$.  Therefore, the joint probability
$\calP_2(\Lam_1, \Lam_2; r)$ and the halo-halo correlation function can
be obtained by considering a large number of realizations.

\subsection{Results}

We present here the results obtained by considering two different
scale-free power-spectra $P(k) \propto k^n$ with $n=-1$ and $n=-2$ in
an Einstein-de Sitter universe.  In these cases the evolution of
clustering is self-similar and the results obtained at a particular
epoch are representative of the whole history.  These two values of the
spectral index can be thought as typical of any physically reliable
power-spectrum on scales relevant for galaxy formation in a
hierarchical scenario.  Adopting a standard procedure we normalize the
power-spectrum so that the linear mass variance as measured in $8 \hm$
spheres is equal to 1 and we impose $\de_c=1.686$.

Concerning the selection of the mass ranges that identify different
classes of haloes, we optimized their broadness in order to balance the
CPU time requirement (too narrow ranges turn out to be poorly
statistically populated) with an accurate description of clustering.
For these reasons we selected three different classes of objects for
each power-spectrum (the first contains objects with $M\gg M_\ast$, the
second has $M \simeq M_\ast$ and the third $M \ll M_\ast$) by requiring
that they are roughly equi-populated (in terms of first upcrossing
events, not number of haloes).  In this way we can explore all the
expected regimes of Lagrangian clustering (a biased halo distribution
for $M \gg M_\ast$, an almost unclustered distribution for $M \simeq
M_\ast$ and an antibiased distribution for $M\ll M_\ast$) with
approximately equal numerical accuracy.

Table 1 gives the parameters that define the three classes of haloes we
selected and the probability of occurrence of first upcrossing events
in each of them [this is obtained by integrating eq. (\ref{Smir}) over
the corresponding interval of $\Lam$].
\begin{table}
\centering
\caption[]{Parameters that identify the classes of haloes.} 
\tabcolsep 4pt
\begin{tabular}{cccccccc} \\ \\ \hline \hline
Class & $\Lam_{\rm min}$ & $\Lam_{\rm max}$ & $M_{\rm min}/M_\ast$ &
$M_{\rm max}/M_\ast$ & $M_{\rm min}/M_\ast$ & $M_{\rm max}/M_\ast$ &
$\int_{\Lam_{\rm min}}^{\Lam_{\rm max}}\calP_1(\Lam) \; d\Lam $ \\
 & & & ($n=-1$) & ($n=-1$) & ($n=-2$) & ($n=-2$) & \\ 
\hline
$I_1$ & 0.45 & 1.79 & 2 & 16 & 4 & 256 & 0.20 \\
$I_2$ & 1.79 & 4.51 & 1/2 & 2 & 1/4 & 4 & 0.22 \\
$I_3$ & 4.51 & 11.37 & 1/8 & 1/2 & 1/64 & 1/4 & 0.19 \\    
\hline
\end{tabular}
\label{tabella}
\end{table}
A technical problem one has to deal with is the occurrence of spurious
oscillations in the correlation function induced by the sharp $k$-space
filter.  In fact, the sharpness of the smoothing kernel unavoidably
gives rise to oscillations in the mass correlations computed in the
Fourier-conjugate space.  To exemplify, by convolving the linear
density field with $\fW_{SKS}(k, R_f)$ one obtains $\xi_m(r;\Lam)
\propto [1-\cos(k_f(\Lam)r)]/r^2$ for $n=-1$ and $\xi_m(r;\Lam) \propto
{\rm Si(k_f(\Lam)r)}/r$ for $n=-2$, where ${\rm Si}(z)\equiv\int_0^z
dy\, j_0(y)$.  Unfortunately, this oscillating behaviour affects also
$\xi^{hh}$.  This can be easily understood in the Langevin formalism,
where the oscillations are generated by the unsmoothed Bessel function
appearing in eq. (\ref{lange2xi}), which is unavoidable in the Wiener
process approach.  One can try to overcome this problem by replacing in
eq. (\ref{DE}) the sharp $k$-space filtered correlation $\xi(r;\Lam)$
with the one obtained with top-hat smoothing.  This is analogous to the
technique generally applied to compare the mass function predicted by
the excursion set approach to the outputs of N-body simulations
(e.g. Lacey \& Cole 1993, 1994). This method strongly reduces the
oscillations but, for $n>-2$, is not able to erase them completely at
separations comparable to the halo size.  In fact, for $n=-1$ and
top-hat filtering, the term $a_{21}$ defined in eq. (\ref{DE}) is
positive for $\Lam ~\mincir \sigma^2(r)$, becomes negative for $\Lam
~\magcir \sigma^2(r)$ and rapidly approaches zero for $\Lam \gg
\sigma^2(r)$, after assuming a minimum negative value.  This term plays
a fundamental role in the computation of the halo correlation function,
causing the appearance of oscillations when $r$ is comparable to the
Lagrangian radius of the given halo.  This problem could be totally
avoided by coherently adopting from the beginning a more realistic
(i.e. non-sharp $k$-space) window function, as sketched in Appendix
B. The price one had to pay, however, is that of dealing with a
space-correlated set of coloured stochastic processes.  In the
following we will always use eq. (\ref{DE}) and top-hat smoothing to
obtain $\xi^{hh}$.

In order to compute the correlation function, for each physical
separation $r$ we followed the evolution of many realizations of the
stochastic processes $\de_1$ and $\de_2$, until $10^6$ pairs of
trajectories crossed the threshold at resolutions $\Lam_1$ and
$\Lam_2$, both contained in one of the three selected intervals.  The
lag $r$ is taken in the range $1\leq r/R_\ast<12 $ for $n=-1$ and
$1\leq r/R_\ast<40$ for $n=-2$, where $R_\ast$ is the Lagrangian radius
associated to the characteristic halo mass $M_\ast$.

Each simulation has been repeated several times (20 for $n=-1$ and 8
for $n=-2$) using different sequences of pseudo-random numbers to build
up the trajectories.  The values for $\xi^{hh}(r)$ shown in Fig. 3 are
obtained by averaging over the different simulations while the error
bars represent the standard deviation of the mean.

In Fig. 3 we also plot the analytic expressions predicted by two
different models: our approximated solution of the Fokker-Planck
equation in eq. (\ref{appp2}) [see also eq. (A1)] and the `counting
field' model described in eq. (14) of CLMP [see also eq. (A2)].  Let us
recall that these two models give rise to the same Lagrangian bias
factors, i.e. to the same clustering regime, provided the halo
separation is a few times larger than their Lagrangian size.
Asymptotically, both the two analytic forms and the numerical
integrations tend to the lowest non-vanishing term of the series
expansion in eq. (\ref{serie}), which, except for the mass range
centered on $M_\ast$, coincides with the Mo \& White (1996) prediction.
In general, the agreement of both models with the results of the Monte
Carlo simulations is remarkably good, except for lags of order the halo
Lagrangian size.  Here the discrepancy between analytical and numerical
results becomes larger and larger as the ratio $M/M_\ast$ decreases.
In particular, none of the two models is able to follow the detailed
features of the numerical solution at small separations, where, at
least for $n=-1$, the spurious oscillations induced by the adoption of
top-hat smoothing in eq. (\ref{DE}) -- which has been instead derived
after sharp $k$-space filtering -- play a relevant role.

We can try to give a more detailed description of the numerical
outcomes by introducing in eq. (14) of CLMP [which is reported
explicitly in Appendix A, eq. (A2)] an extra modulation induced by a
decaying sinusoidal term. Let us call $\xi_{CLMP}$ the analytic form
for $\xi^{hh}$ appearing in eq. (14) of CLMP. We can then introduce the
following `best-fitting models'
\be
\f{\xi^{hh}(r)-\xi_{CLMP}(r)}{1+\xi_{CLMP}(r)}=
C_1 \cos \left(C_2 \f{r}{R_\ast}+C_3\right) \exp \left[-C_4 
\left( \f{r}{R_\ast}\right)^2 \right] \;,
\label{fit1}
\ee
\be
\f{\xi^{hh}(r)-\xi_{CLMP}(r)}{1+\xi_{CLMP}(r)}=
C_1 \cos \left(C_2 \f{r}{R_\ast}+C_3\right)\exp\left[-C_4 \f{r}{R_\ast}
\right] \;,
\label{fit2}
\ee
respectively for $n=-1$ and $n=-2$.  The coefficients $C_\alpha$
($\alpha=1,\dots,4$), which are found using the Levenberg-Marquardt
non-linear least-squares method (e.g. Press \etal 1992) in each mass
range, are given in Table 2.  The main limitation of this `best-fit'
approach is clearly that the coefficients $C_\alpha$ depend both on the
shape of the power-spectrum and on the halo masses.  Although we did
not attempt any such approach, it ought to be possible to parametrize
these two dependences.
\begin{table}
\centering
\caption[]{Parameters for the `best-fitting models'. In the last
row the value of $C_3$ is set to $\pi$ since $C_2\simeq 0$
causes a degeneracy between $C_1$ and $\cos(C_3)$.
This is removed by assigning an arbitrary value to $C_3$.}
\tabcolsep 4pt
\begin{tabular}{cccccc} \\ \\ \hline \hline
$n$ & Class & $C_1$ & $C_2$ & $C_3$ & $C_4$  \\ \hline
$-1$ & $I_1$ & $0.24 \pm 0.02 $ & $2.47\pm0.16$ & $-2.7\pm 0.3 $ &
$0.570 \pm 0.016 $ \\  
$-1$ & $I_2$ & $1.69 \pm 0.10 $ & $1.34\pm0.07$ & $0.00\pm 0.08 $ &
$0.880 \pm 0.008 $ \\ 
$-1$ & $I_3$ & $21 \pm 6 $ & $ 0.65\pm0.13$ & $1.1\pm 0.2 $ &
$2.49 \pm 0.13 $ \\ 
\hline
$-2$ & $I_1$ & $7.0 \pm 1.5 $ & $0.35\pm0.07$ & $6.8\pm 0.2 $ &
$1.050 \pm 0.012 $ \\  
$-2$ & $I_2$ & $57 \pm 11 $ & $0.08\pm0.16$ & $1.49\pm 0.14 $ &
$1.90 \pm 0.02 $ \\ 
$-2$ & $I_3$ & $0.86 \pm 0.12 $ & $ 0.0 \pm 0.2$ & $ \pi$ &
$1.475 \pm 0.011 $ \\
\hline
\end{tabular}
\label{tabella}
\end{table}
\begin{figure}
\centering{
\vbox{
\hbox{
\psfig{figure=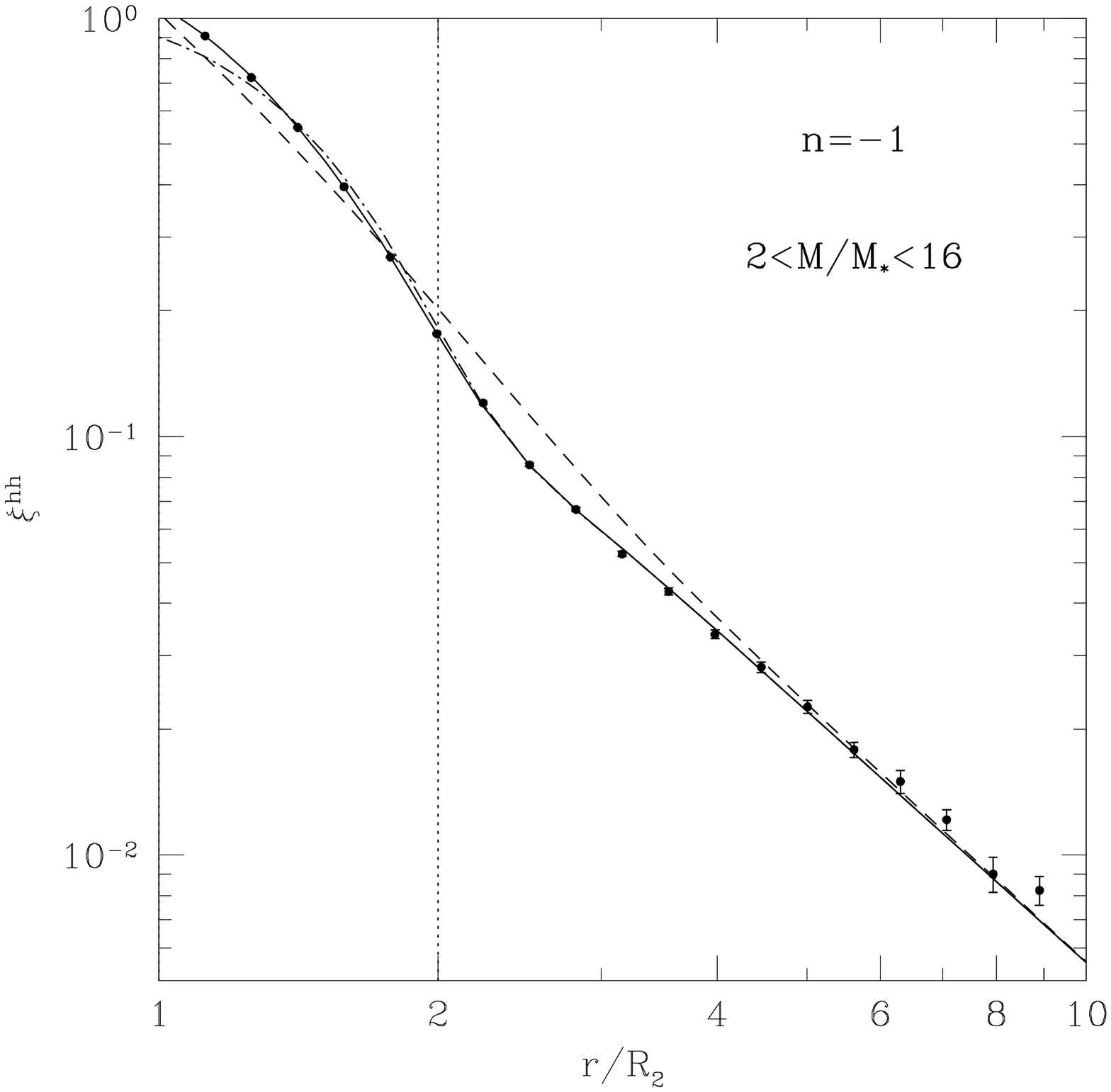,bburx=574pt,bbury=695pt,clip=t,width=7cm}
\psfig{figure=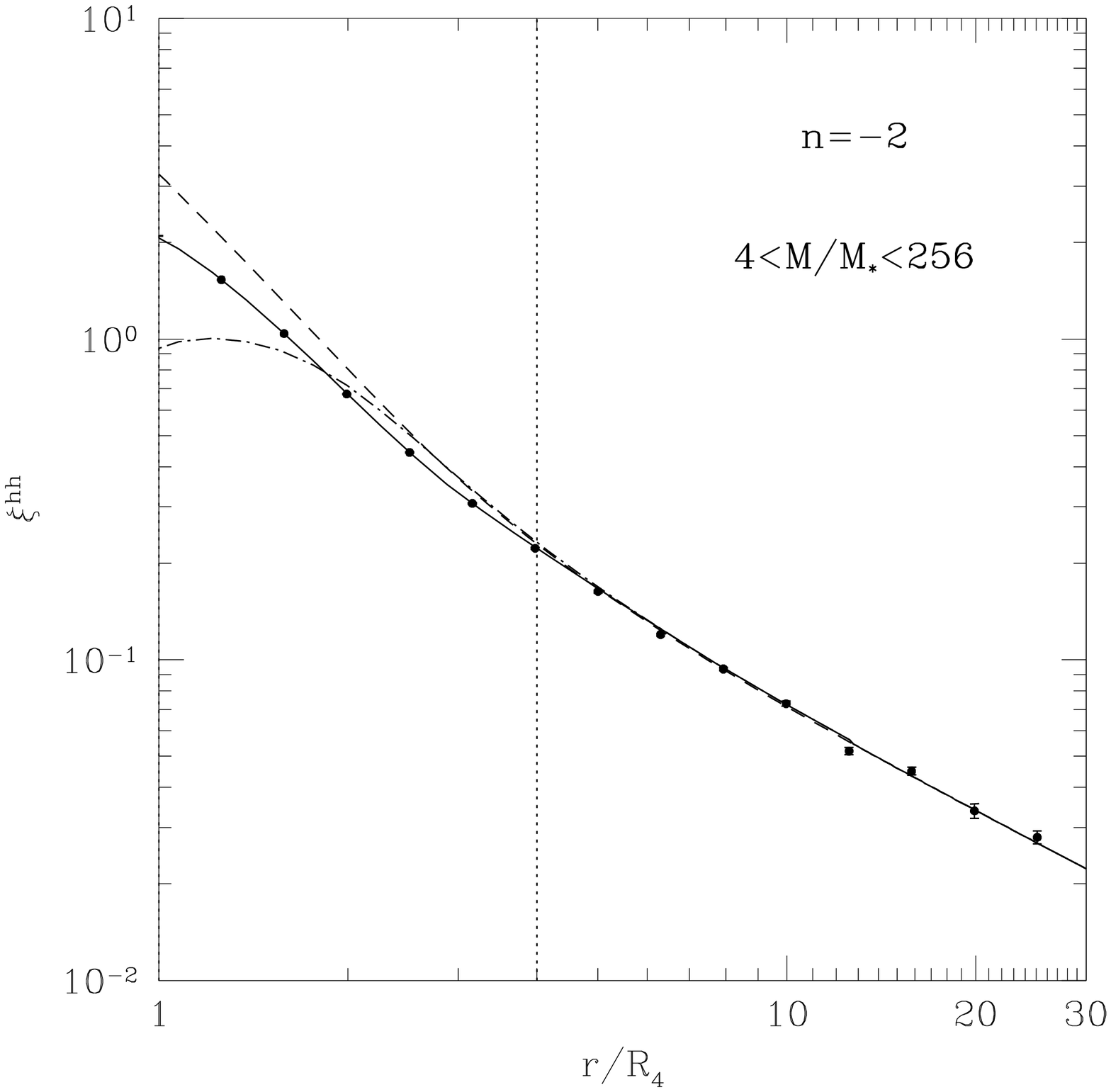,bburx=574pt,bbury=695pt,clip=t,width=7cm}
}
\hbox{
\psfig{figure=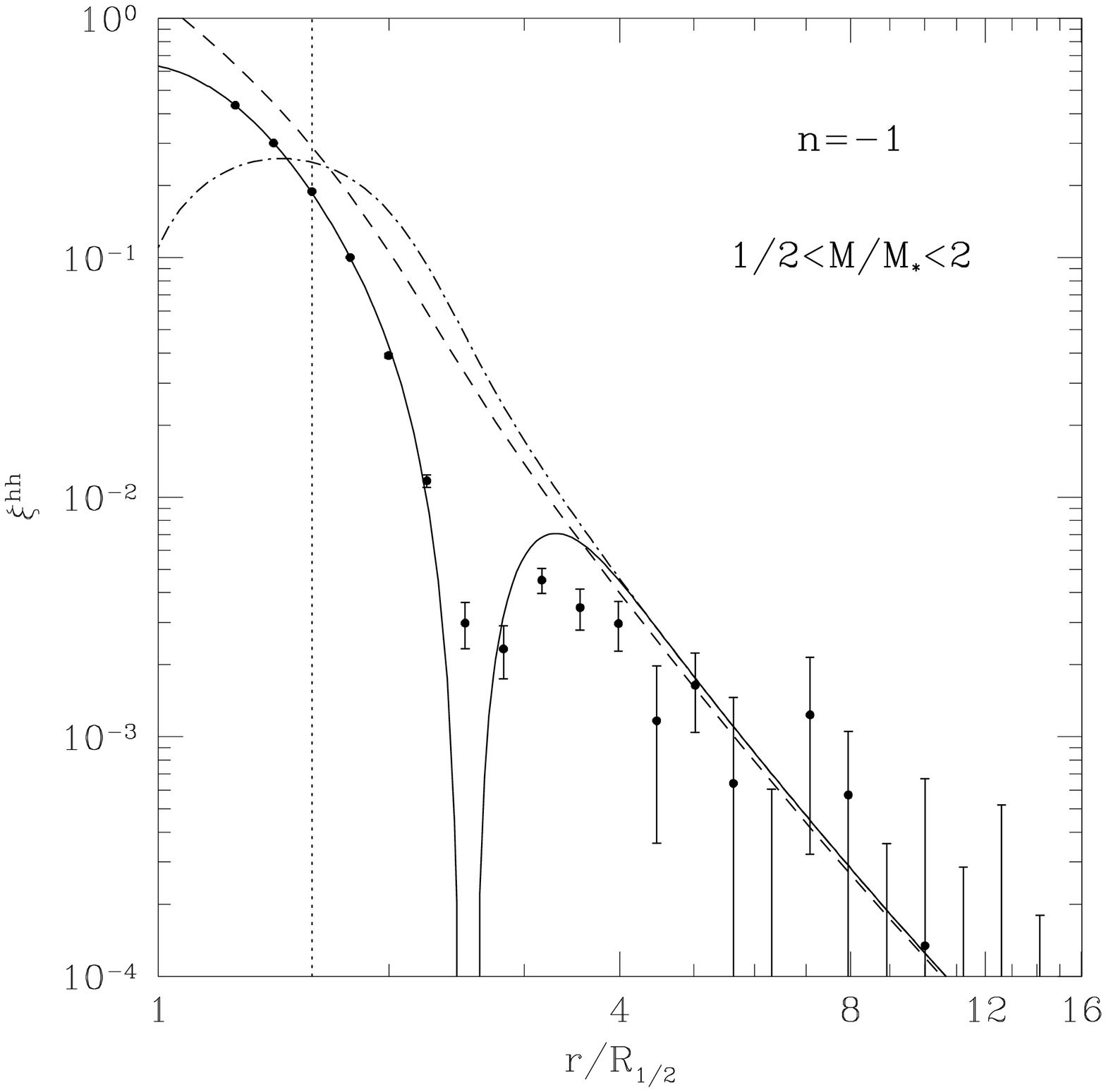,bburx=574pt,bbury=695pt,clip=t,width=7cm}
\psfig{figure=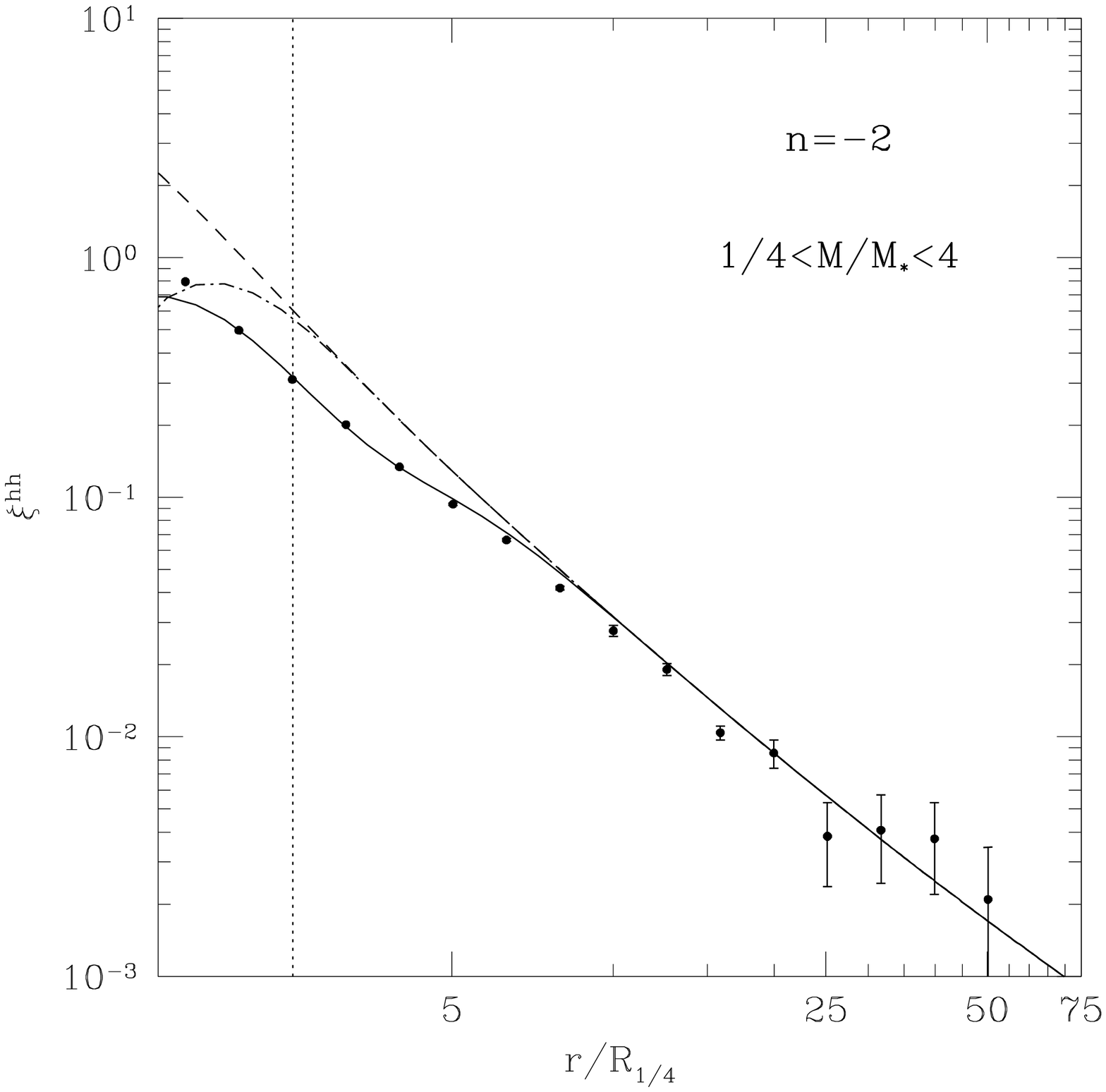,bburx=574pt,bbury=695pt,clip=t,width=7cm}
}
\hbox{
\psfig{figure=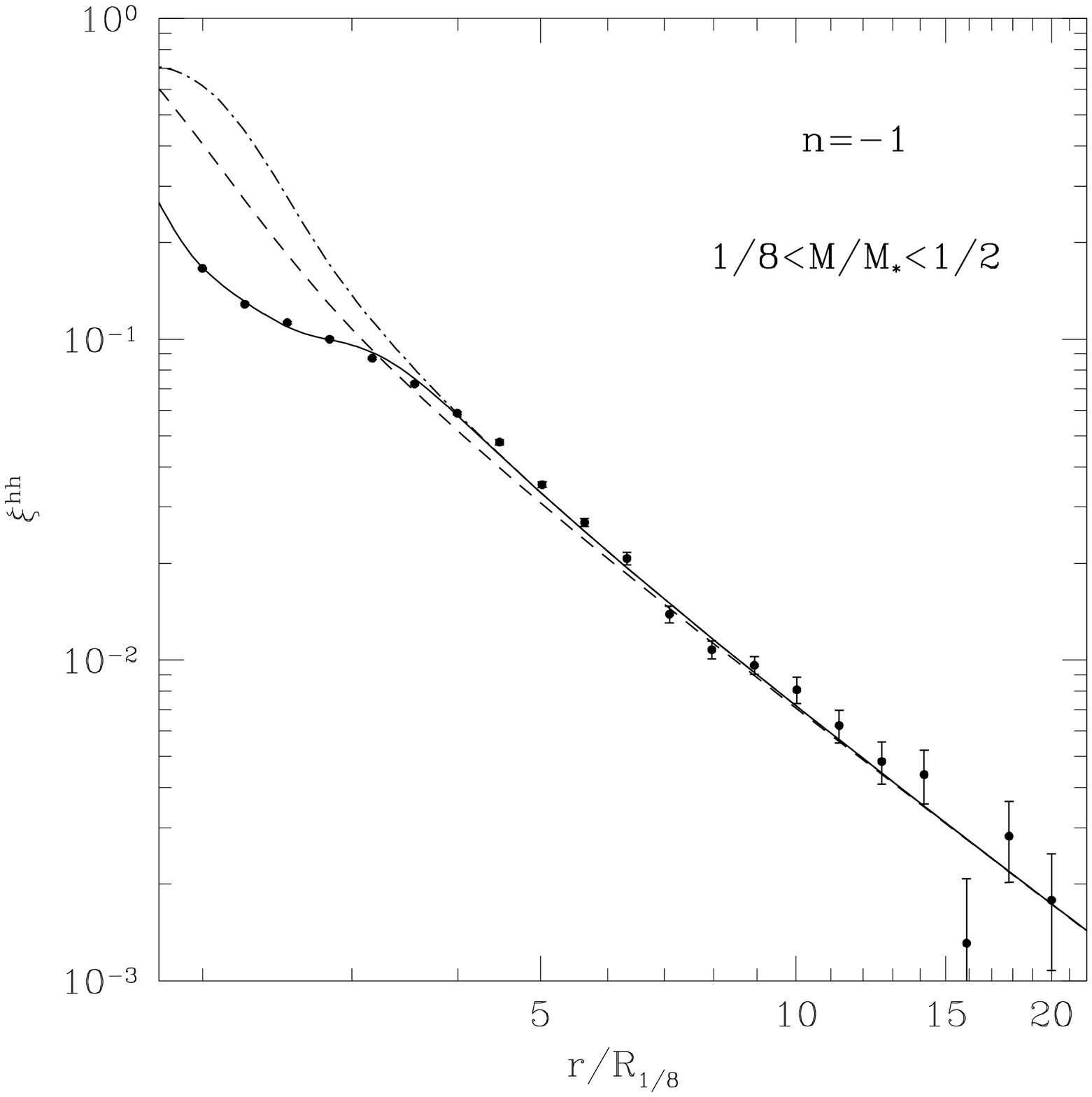,bburx=574pt,bbury=695pt,clip=t,width=7cm}
\psfig{figure=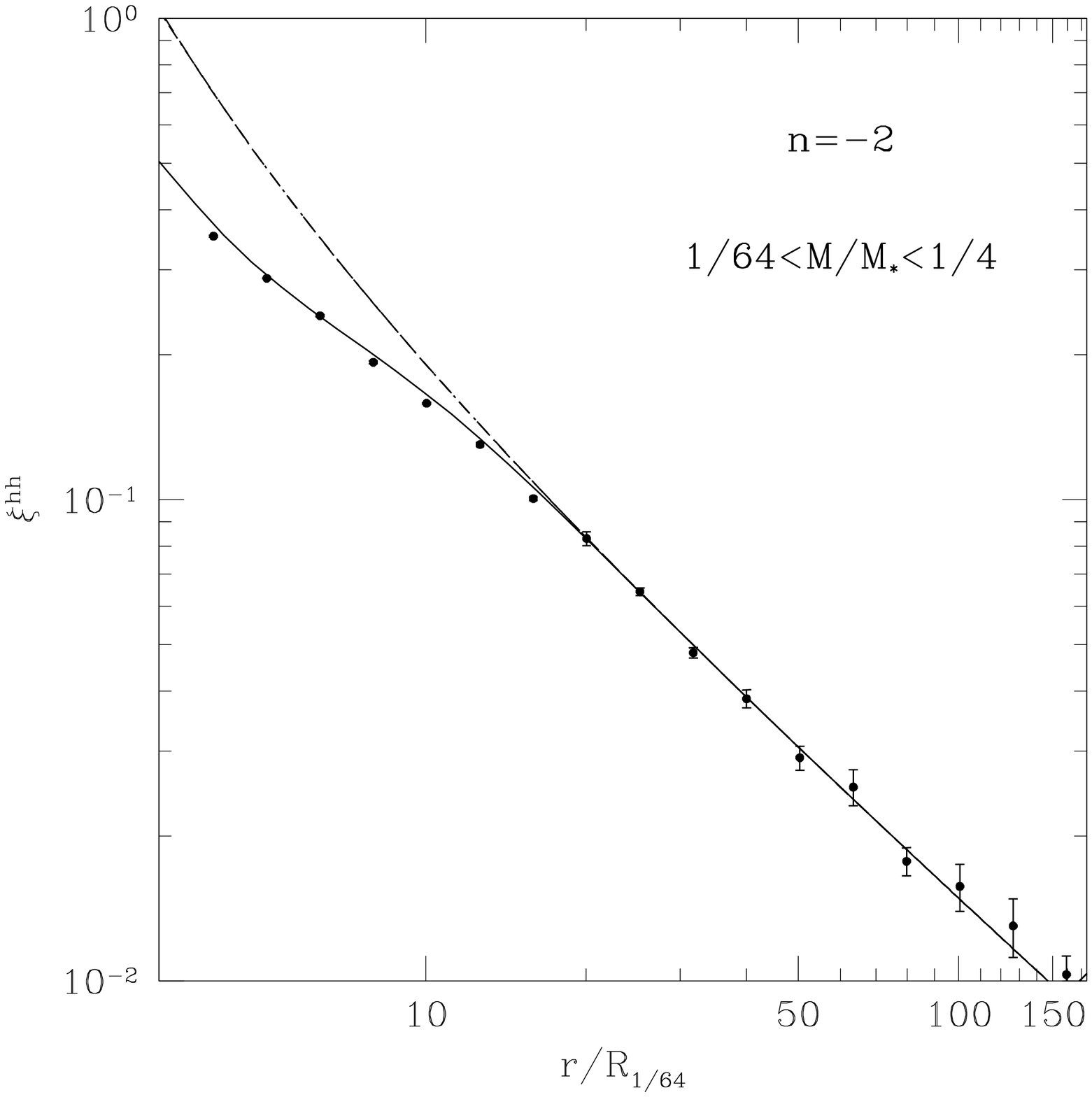,bburx=574pt,bbury=695pt,clip=t,width=7cm}
}
}
}
\caption{The Lagrangian halo correlation function $\xi^{hh}$ in an
Einstein-de Sitter universe with two different scale-free
power-spectra, $n=-1$ ({\em left} column) and $n=-2$ ({\em right}
column), is shown for three halo mass ranges. The object separation $r$
is scaled to the Lagrangian radius of the least massive halo in each
range. The vertical dotted lines, where shown, are placed at the
Lagrangian radius of the most massive halo in each range.  The points
represent the mean value of different realizations obtained by
numerically solving our correlated Langevin equations, while the error
bars represent the scatter of the mean.  The {\em continuous} lines
refer to the `best-fit models' of eqs. (\ref{fit1}) and
(\ref{fit2}). The {\em dashed} lines are obtained from our approximated
solution of the Fokker-Planck equation in eq. (\ref{appp2}) [see also
eq. (A1)], while the {\em dot-dashed} lines show the predictions of the
`counting field' model of eq. (14) in CLMP [see also eq. (A2)].
Top-hat filtering is used in all cases.}
\label{fig:3}
\end{figure}
%
%

\section{Conclusions}

In this paper we have proposed a model for the clustering of dark
matter haloes in Lagrangian space.  Our model is based on a natural
extension of the excursion set approach to the PS theory (e.g. BCEK),
namely we accounted for the spatial correlations of the linear mass
density fluctuation field.

In particular, the two-point halo correlation function has been
obtained by integrating the system of Langevin equations governing the
evolution of pairs of correlated density processes or, equivalently for
a Gaussian mass distribution, the bivariate Fokker-Planck equation for
the density probability distribution of the same processes, once
appropriate boundary conditions have been imposed. We believe that the
numerical integration of the correlated Langevin equations allowed for
the most reliable determination of the Lagrangian halo correlation,
complementing the Press-Schechter inspired analyses of the halo mass
function.

Although we gave explicit results only for the halo two-point function,
a generalization to higher order statistics would be
straightforward. The halo correlation function obtained with the
present approach is fully consistent with the recent results obtained
by CLMP, and its form at large separation reproduces the linear bias
relation by Mo \& White (1996), which has been shown to be in good
agreement with the clustering of synthetic haloes in N-body
simulations.

As stressed by CLMP, the issue of transforming the halo distribution
from the Lagrangian to the Eulerian world is a non-trivial one,
especially on smaller scales and for smaller mass systems, which are
most affected by the intrinsic non-linearity of the evolved mass
density field and by the occurrence of multi-streaming. Nevertheless,
one might speculate that a stochastic approach analogous to the one
considered so far would be viable also in Eulerian space.  The main
modification induced by the Lagrangian-to-Eulerian map would in fact be
a {\em local} modulation of the halo formation threshold, namely $
t_f(z_f) \longrightarrow t_f'(\bx ,z_f) \equiv t_f(z_f) -
\de_{M_\0}(\bx ,z_f)\;, $ with $\de_{M_\0}(\bx ,z_f)$ the Eulerian mass
density contrast smoothed on some background mass scale $M_\0$ much
larger than the halo one.

The results of this paper represent a relevant step towards the
construction of a local identification algorithm for halo formation
sites in Lagrangian space, which would allow to depict halo maps
starting from low-resolution simulations of the evolved dark matter
distribution. As a straightforward application of this general idea one
can apply the Zel'dovich approximation (Zel'dovich 1970) to follow the
dark matter dynamics on mildly non-linear scales and use it to
reconstruct the local halo number density at various epochs (Catelan,
Matarrese \& Porciani 1998).

One of the most interesting direct applications of our technique is the
construction of spatially correlated halo merging
trees. State-of-the-art algorithms to follow mass accretion histories
are entirely based either on the mean one-point distribution of the
first upcrossing events $\calP_1$ (Lacey \& Cole 1993; Kauffmann \&
White 1993; Somerville \& Kolatt 1997) or on Monte Carlo realizations
of the one-point Wiener process $\delta(\Lam)$ (e.g. Tozzi \etal 1996).
In both cases one has to extrapolate the succession of merging events
involving a large number of haloes coming from different Lagrangian
regions from the knowledge of the average properties of a single mass
accretion history. Obviously some {\em ad hoc} assumptions need to be
made in the tree reconstruction process, to supply for the lack of
statistical and spatial information; examples are the assumption of
binary merging in Lacey \& Cole (1993) or the recent efforts aimed at
distinguishing between mass accretion and merging events
(Salvador-Sol\'e \etal 1997; Somerville \& Kolatt 1997).

In a more realistic approach, however, each branch of a merger tree
should be associated to a different Lagrangian position. Thus, spatial
correlations of the initial density field would manifest themselves as
statistical correlations both between different branches of the same
tree and between different trees. The importance of this issue was
stressed already in Lacey \& Cole (1993). Somerville \& Kolatt (1997)
argued that accounting for the complex correlated structure of the
density field would permit the construction of merger trees starting at
high redshift and propagating forward in time.

A first attempt to include spatial correlations in modeling the
hierarchical growth of dark matter haloes was made by Yano, Nagashima
\& Gouda (1996) and by Rodrigues \& Thomas (1996).  The main effort of
these authors was however devoted at solving the cloud-in-cloud problem
when the Lagrangian regions that ultimately collapse into haloes are
properly treated as extended ones.  Their predictions for the mass
function imply more high-mass objects and less low-mass haloes than the
PS expression.

However, the issue we want to focus on is a very different and more
ambitious one. By adopting the excursion set theory, we aim at building
realizations of merger histories by following the fate of the
correlated trajectories associated to each branch of the same tree.
Since we are not modifying the one-point distribution of first
upcrossing events compared with the standard solution of the
cloud-in-cloud problem (BCEK) our merging histories are {\em a priori}
consistent with the PS mass function.

Concerning other quantities that characterize the ensemble of merging
histories (such as halo survival and formation times) we expect that
accounting for spatial correlations will lead to relevant
modifications. In general, compared with the uncorrelated case, we
should obtain corrections whose importance depends on the mass and the
epoch under consideration.  Intuitively, we would expect the larger
effects on those objects whose mass accretion histories are dominated
by merging of halos with similar masses.  In fact, as shown in
Fig. \ref{trajectories}, the correlation between nearby trajectories is
important only when the smoothing length is larger than or comparable
to the physical separation between the Lagrangian points to which the
trajectories are associated.  In a merger tree the physical separation
between two branches should be roughly given by the sum of the
Lagrangian radii of the haloes measured just before they merge.
Therefore, the effect of spatial correlations of the density field will
be completely negligible if the two Lagrangian halo sizes are much
different.  These and related issues will be analyzed in a forthcoming
paper.

\section*{Acknowledgments}

CP is grateful to Riccardo Mannella for useful discussions on
state-of-the-art numerical simulations of stochastic processes and in
particular for focusing our attention on the 1987 unpublished paper by
W. Strittmatter, which gave us the idea of using an intra-step Monte
Carlo simulation to correct for time-discreteness. PC has been
supported by the Danish National Research Foundation at the Theoretical
Astrophysics Center in Copenhagen and by the EEC at the Department of
Astrophysics in Oxford. PC is grateful to George Efstathiou and Cedric
Lacey. FL, SM and CP thank the Italian MURST for partial financial
support.

\appendix

\section{Analytic expressions for the halo two-point function}

For the sake of completeness we give here the full expression of two
approximations for $\xi^{hh}$ frequently referred to in the main text
and used in Fig. \ref{fig:3}.  These formulae represent the results of
different generalizations of the PS formalism, either in the
probabilistic Fokker-Planck formulation of the excursion set approach,
discussed in Section 3.2.3, or in the local counting field theory given
by CLMP.

The explicit result for the halo correlation deriving from the ansatz
introduced in Section 3.2 is obtained by inserting eqs. (\ref{appp2})
and (\ref{Smir}) in eq. (\ref{appsol}).  Thus, for the cross
correlation between haloes selected in the infinitesimal mass ranges
corresponding to the intervals $\Lam_1+d\Lam_1$ and $\Lam_2+d\Lam_2$,
one gets
\ba
1+\xi^{hh}(r) &=&
\f{t_f^2 \Lambda_1 \Lambda_2+\left[
 \Lambda_1 \Lambda_2 -t_f^2(\Lambda_1+\Lambda_2) \right]
\xi(r;\Lam_{\rm m})+t_f^2 \,\xi(r;\Lam_{\rm m})^2-\xi(r;\Lam_{\rm m})^3}
{\Lam_1^{-3/2} \Lam_2^{-3/2} \,
\left[\Lambda_1 \Lambda_2 -\xi(r;\Lam_{\rm m})^2 \right]^{5/2}} \times 
\nonumber \\
&\times&
\exp{\left[-\f{t_f^2}{2}~ \f{\left(\Lam_1+\Lam_2 \right)
\xi(r;\Lam_{\rm m})^2-2 \,\Lam_1 \Lam_2 \,\xi(r;\Lam_{\rm m})}
{\Lam_1 \Lam_2 \left[\Lambda_1 \Lambda_2 -\xi(r;\Lam_{\rm m})^2\right]}
\right]}\;.
\label{def1}
\ea
On the other hand, the solution in the CLMP model, once their eq.(14)
has been recast in the present notation, is
\ba
1+\xi^{hh}_{CLMP}(r)
\, &=&\,\f{1}{\sqrt{1-\w^2}}
\left\{
1+\f{\s_2^2}{(1-\om^2)}
\,\left( \f{1}{\s_1}- \f{\om}{\s_2} \right)
\,\f{\p\om}{\p \s_2} \, \right. 
+ \,\f{\s_1^2}{(1-\om^2)} \, 
\left(\f{1}{\s_2}-\f{\om}{\s_1}\right) 
\,\f{\p\om}{\p \s_1}\,+ 
\,\f{\s_1^2 \s_2^2}{t_f^2} \, 
\f{\p^2\om}{\p \s_1\p \s_2} 
\nonumber \\ 
&+&
\f{\s_1^2\s_2^2}{t_f^2 (1-\om^2)^2}\, 
\left[\w(1-\w^2)+
(1+\om^2)\,\f{t_f^2}{\s_1 \s_2} \,-\, 
\om\,t_f^2\,\Big(\f{1}{\s_1^2}+\f{1}{\s_2^2}\Big)\right] 
\f{\p\om}{\p {\s_1}}\,\f{\p \om}{\p {\s_2}} \bigg\} \,
\nonumber \\
&\times&\exp {\Bigg[ - \displaystyle{ \f{t_f^2}{2}} \,
\displaystyle{ \f{ \om^2 
\Big( \displaystyle{
\f{1}{\s_1^2} } + \displaystyle{ \f{1}{\s_2^2}} 
\Big) -
\displaystyle{ 2\,\f{\om}{\s_1 \s_2} }} 
{\left( 1- \om^2 \right)}} \Bigg]}\;.
\label{eq:k1}
\ea 
Here $\s_i=\Lam_i^{1/2}$ and $\w=\xi(\Lam_1, \Lam_2; r)/\s_1 \s_2$ with
$\xi(\Lam_1, \Lam_2; r)$ the correlation function between the linear
mass density fluctuation field smoothed with two different resolutions
$\Lam_1$ and $\Lam_2$.  For sharp $k$-space filtering $\xi(r;\Lam_1,
\Lam_2)= \xi(r;\Lam_{\rm m})$.  It is worth mentioning that for
separations larger than the smoothing lengths, when the derivatives of
$\xi(r;\Lam_1, \Lam_2)$ with respect to $\Lam_1$ and $\Lam_2$ are
negligible, eq. (\ref{eq:k1}) reduces to eq. (\ref{def1}).  Moreover,
both approximations asymptotically reach the linear bias regime studied
by Mo \& White (1996).

\section{Correlated Langevin equations: general smoothing}

Collecting the results outlined in Section 2.2, we can describe the
evolution of an ensemble of pairs of correlated trajectories by solving
the system of Langevin equations
\be 
\left\{
\begin{array}{l}
{\displaystyle {d \de_1(R_f) \over  dR_f}=\eta_1(R_f)\;,
\;\;\;\;\;\;\;\;\;\;\;\;\;\;\;\;\;   
\lim_{R_f \to \infty}\delta_1(R_f)=0 }\;, \\ \\
{\displaystyle {d \de_2(R_f) \over  dR_f}=\eta_2(R_f)\;,
\;\;\;\;\;\;\;\;\;\;\;\;\;\;\;\;\;
\lim_{R_f \to \infty}\delta_2(R_f)=0 } \;, \\ \\
\langle \eta _1(R_f) \rangle=\langle \eta _2(R_f) \rangle=0\;,
\;\;\;\;\;\;\;\;
\eta_1\;\,{\rm and}\;\, \eta_2\;\,
{\rm Gaussian\;\;\, processes}\;, \\ \\
{\displaystyle  \langle \eta_1(R_f)\, \eta_1(R'_f)\rangle =
\langle \eta_2(R_f)\, \eta_2(R'_f)\rangle =
{1\over 2\pi^2} \int_0^\infty dk\, k^2\, P(k)\, 
{\p \fW(kR_f)\over \p R_f}\,
{\p \fW(kR'_f) \over \p R'_f}}\;,
 \\ \\
{\displaystyle \langle \eta_1(R_f) \,\eta _2(R_f') \rangle =
{1\over 2\pi^2} \int_0^\infty dk\, k^2\, P(k)\, 
{\p \fW(kR_f)\over \p R_f}\,
{\p \fW(kR'_f) \over \p R'_f}\,
j_0(kr)\;.}
\end{array}
\right.
\label{lange2eta}
\ee
We wrote here the equations adopting the same notation as in Section
2.2, i.e.  using the smoothing radius $R_f$ as time variable for the
trajectories.  In this way we emphasize the role played by the filter
in determining the statistical properties along and between the
trajectories.  For sharp $k$-space filtering, the quantity $\p
\fW(kR_f)/\p R_f$ reduces to a Dirac delta function, which leads to the
set (\ref{lange2xi}) of the main text.  On the other hand, the
advantage of a more realistic filter is that the oscillations induced
by the zeroth order Bessel function are damped.

\end{document}